\documentclass[aps,superscriptaddress,amsmath,showpacs]{revtex4-2}
\usepackage{slashed,bbm,hyperref,amsmath,amssymb,amsfonts,graphicx,array}
\usepackage{lineno}
\graphicspath{{Figures/}}
\usepackage{float,subfig,xcolor}
\usepackage[english]{babel}
\usepackage{bigstrut}
\usepackage{multirow,adjustbox}

\newcommand{\calh}{\mathcal{H}}
\newcommand{\beq}{\begin{equation}}
\newcommand{\eeq}{\end{equation}}
\newcommand{\bea}{\begin{eqnarray}}
\newcommand{\eea}{\end{eqnarray}}
\newcommand{\til}{\tilde}
\newcommand{\rhot}{\tilde\rho}

\newcommand{\dg}{\dagger}

\newcommand{\bsplit}{\begin{split}}

\newcommand{\tr}{\mathrm{Tr}}

\begin{document}

\title{The effect of nonequilibrium entropy production on the quantum Fisher information and correlations}

\author{Xuanhua Wang}
\affiliation{Department of Physics and Astronomy, Stony Brook University, SUNY, Stony Brook, NY 11794 USA}
\author{Jin Wang}
\email[]{Corresponding author: jin.wang1@stonybrook.edu}
\affiliation{Department of Physics and Astronomy, Stony Brook University, SUNY, Stony Brook, NY 11794 USA}
\affiliation{Department of Chemistry, Stony Brook University, SUNY, Stony Brook, NY 11794 USA}

\begin{abstract}
In this study, we apply quantum master equations beyond secular approximation, and investigate the nonequilibrium thermodynamic cost of enhanced quantum metrology and quantum correlations. We find that the nonequilibrium conditions enhance quantum Fisher information (QFI) and quantum correlations predominantly for weak tunneling scenarios. The enhancement is assisted by a corresponding increase of the thermodynamic cost characterized by the entropy production rate (EPR). For the strong tunneling regimes, the QFI and quantum correlations can not be unceasingly boosted by higher thermodynamic costs and decay once the system is overburdened with extremely large energy currents. The result indicates that for open systems with weak tunneling rates, thermodynamic cost can be potentially exploited to improve the quantum metrology and quantum correlations.
\end{abstract}

\pacs{03.65.Ud, 03.65.Yz, 03.67.-a, 05.70.Ln}
\maketitle

\numberwithin{equation}{section}
\section{Introduction and motivation}
Quantum enhanced measurement is one of the technologies that has seen tremendous developments during the past decade \cite{nolan2016quantum, cooper2012robust, giovannetti2004quantum, wolfgramm2013entanglement, yonezawa2012quantum}. As early as 1980s, detailed proposals had been put forward to amplify the sensitivity of gravitational wave detection using squeezed light \cite{braun2018quantum}. Similar method was also proposed in the field of spectroscopy, for example, to enhance the fermion interferometer sensitivity \cite{yurke1986input}. One of the standard tools to evaluate the quantum enhancement is the celebrated quantum Cram\'er-Rao bound \cite{scheffe1947h}. The Cram\'er-Rao bound gives the minimum achievable statistical uncertainty in the estimation of the parameters, which is defined as the quantum Fisher information (QFI) \cite{braunstein1994statistical,uhlmann1991gauge}. QFI is a central concept in quantum estimation theory, and it is found instrumental in characterization of the disparity in quantum states \cite{braunstein1994statistical,luo2004informational}. For example, QFI as a measure of the geometric distance between quantum states, has been applied to study quantum phase transitions without clear order parameters \cite{marzolino2017fisher}. In addition, some recent studies have suggested a strong connection between the QFI, the perturbative gravitational energy and the bulk entanglement in holography \cite{banerjee2018connecting,sarkar2017holographic,lashkari2016canonical}.

Quantum entanglement is widely used as a measure of nonclassicality for quantum systems. However, unentangled states can also exhibit non-classical behavior especially when dealing with mixed states \cite{braun2018quantum}. This nonclassical correlation can be captured by quantum discord and mutual information which are used as characterizations of quantum correlations in information theory, quantum computing and also biophysics due to their robustness against the noises from the environments \cite{ollivier2001quantum}. The interplay between the quantum correlations and QFI is a fundamental issue that is pertinent to a wide range of researches. One of the distinct traits that separate a quantum system from its classical counterpart is the ``spooky action at distance'' or quantum entanglement \cite{einstein1935can}, and it is shown to be a resource for enhanced quantum metrology \cite{nakano2013negativity, chitambar2019quantum}. Under certain constraints, the greatest precision in the estimation of a given parameter is constrained by the Standard Quantum Limit (SQL). Nevertheless, by exploiting the very quantum nature of the system, typically by using the entangled input states, a different limit can be saturated which is called the Heisenberg limit (HL) \cite{haase2016precision,toth2014quantum}. Genuine multiparticle entanglement is a crucial requirement for reaching the highest sensitivities, for example quantum, in the study of the interferometry \cite{hyllus2012fisher, braun2018quantum, giovannetti2004quantum, giovannetti2011advances, wiseman2009quantum}. While on the other hand, a large QFI often implies a nonvanishing Bell correlation or other quantum correlations such as quantum discord \cite{frowis2019does, kim2018characterizing}. Despite the close relationship between QFI and quantum correlations as suggested by accumulating studies, a general and conclusive statement remains elusive. Some recent studies questioned that quantum metrology may not be related to the fundamental concept of entanglement and a general model-independent relationship between the entanglement measures and QFI is still yet to be found \cite{przysikezna2015quantum,frowis2019does}.

The study of quantum metrology and the quantum correlations in ambient environment is imperative to fully appreciate the quantum features of complex systems. It has a spectrum of applications to experiments and technologies of thermometry, magnetometry and biophysical systems \cite{spietz2003primary,savukov2005tunable,marzolino2014quantum, wang2018coherence,yao2014quantum}. For example, for quantum information processors (QIP), the interactions with environments limit the fidelity and the quantum enhancement of scaling. This can be detrimental to the precision of measurements and also the entanglement in the system, which is an important factor to consider for QIP design \cite{noiseinf,escher2011general, demkowicz2012elusive,haase2016precision}. Molecular junctions are intrinsically dissipative systems which have been widely studied. Depending on the specific situation, different mechanisms can lead to the tunneling or energy transfer between two carriers such as the F\"oster energy transfer, Dexter energy transfer and surface energy transfer, etc. High precision of estimation for parameters such as the tunneling strength between the two charge carriers is instrumental for the experimental studies of the molecular structure \cite{thoss2018perspective, skourtis2016dexter}. 

In an open system, the nonequilibrium steady state (NESS) has to be sustained by external resources with the consumption of thermodynamic costs. To be specific, a constant current of energy input into the system is required to keep the system in the steady state. This irreversible process is marked by the production of entropy \cite{santos2019role}. Thermodynamic cost is not only central and unavoidable for performing computations which is constrained by the Landauer limit \cite{landauer1961irreversibility}, but also finds useful in enhancing quantum coherence and facilitating quantum measurements \cite{santos2019role,deffner2016quantum}.  

Many studies on dissipative systems are primarily concentrated on the equilibrium scenarios or the relaxation process \cite{troiani2018universal,chin2012quantum}. Our goal is to explore the intrinsic nonequilibrium effect with multiple reservoirs, and quantify the influence of the nonequilibriumness and the thermodynamic costs on the quantum metrology and quantum correlations of the system. Classically, multiple nonidentical environments lead to the appearance of heat or particle flux and the appearance of the thermodynamic cost characterized by the entropy production under the breaking of the detailed balance \cite{fang2019nonequilibrium,de2013non}. Quantum mechanically, this nonequilibriumness was shown to induce the quantum nonlocality and enhance the coherence besides the ``probability flow" \cite{wang2019steady, wang2018coherence, wang2019nonequilibrium, zhang2014curl, li2015steady,zhang2020influence,zhang2021entanglement} and plays an important role in the appearance of strong entanglement which is important for the study of quantum thermal machines \cite{brask2015autonomous,tacchino2018steady}. In particular, for a weakly-interacting bipartite system in the NESS, quantum correlations between the two subsystems are significantly strengthened as the two parts are driven away from the equilibrium \cite{wang2019nonequilibrium}. This NESS coherence which is sustained by the nonequilibrium environments was shown to be resourceful for quantum metrology in quantum systems such as optical molecular models and beneficial for quantum energy transfer and quantum refrigerator in the study of quantum thermal machines and light-harvesting systems \cite{brunner2014entanglement,manzano2012quantum,brask2015autonomous,tacchino2018steady,wang2018coherence,wang2021excitation}.

To address the effect of nonequilibrium environments and the thermodynamic costs, we study a simplified model of molecular junctions, which comprises two interacting fermion carriers immersed in two separate reservoirs. The model and its variations have been applied extensively to study quantum dots, charge transport in many-body system, thermal entanglements, designs of quantum information processors, and especially the study of molecular junctions \cite{onuchic1988classical, sinaysky2008dynamics, gunlycke2001thermal, petrosyan2002scalable, petrosyan2006quantum, joachim2005molecular, galperin2005hysteresis}. The archetypical model of a molecular junction constitutes two metal or semiconductor electrodes (which can be modeled as charge reservoirs), molecules and their mutual coupling. This model presents the possibility to study the plethora of nonequilibrium behavior with the prospect of technological applications in nanoelectronic devices \cite{thoss2018perspective}. The natural parameter in this model is the tunneling strength between the two subsystems. The strength of the intra-system tunneling is finely tuned to dictate the rate of charge and energy transport, and it is often directly related to the distance between the two carriers and is closely related to the study the molecular structures in the biophysics experiments. 

The environment is frequently deemed as a negative contribution in the study of quantum metrology; however, we show that when one has certain control over the environment, it can be used as a booster in the accuracy. This point has not been carefully addressed, for example, from the thermodynamic perspective in terms of entropy production. The related result is new and not discussed in our previous work \cite{wang2019nonequilibrium}. The result is an extension on the effects of nonequilibriumness in quantum metrology. We believe that the model we use here is simple enough and clear enough to convey the idea of nontrivial nonequilibrium impacts on quantum metrology.

In this paper, we apply the Bloch-Redfield equation beyond the secular approximation and quantify the performance of the parameter estimations of the tunneling rate (also termed as the coherent coupling or RDDI potential depending on the specific models) with the QFI and the quantum correlations with nonequilibrium thermodynamic costs. In section two, we introduce the QFI and quantum correlations. In sections three and four, we introduce the system Hamiltonian and we present the approximate analytic solutions to the equations of motion. In section five, we investigate the quantum metrology and quantum correlations in the NESS and how the thermodynamic costs can be exploited to enhance them significantly in a nonequilibrium system in the weak coupling regimes. Finally, we demonstrate the nonmonotonic behaviors of the QFI and correlations in the strong coupling regime and comment on the limitations of the enhancement of the QFI and quantum correlations by escalating the thermodynamic costs. The explicit expressions of the dissipators and the proof of positivity of the EPR semi-classically defined in our paper are given in the Appendix.

\numberwithin{equation}{section}
\section{QFI and quantum correlation measures}
In parameter estimation theory, Fisher information plays gives the upper bound of accuracy of estimations \cite{braun2018quantum,pezze2018quantum}. For a random variable $x$ with probability distribution $p_{\theta}(x)$, where $\theta$ is the parameter of the distributions, we can estimate the parameter $\theta$ based on the observed distribution. However, this estimation may deviate from the real value and the variance of $\theta$ is given by the Cram\'er-Rao (lower) bound $\text{Var}(\theta_{\rm est})\ge \frac{1}{J_\theta^{(M)}}$, where $ \text{Var} (\theta_\text{est}) :=\langle(\theta_\text{est} - \theta)^2\rangle$ is the variance of $\theta_\text{est}$ and the (classical) Fisher information $J_\theta$ is defined as
\begin{equation}
  J_\theta^{(M)}= \int d^M x \frac{1}{p_\theta(x)}\left(\frac{\partial p_\theta(x)}{\partial \theta}\right)^2\,. \label{cfi}
\end{equation}

For a quantum system, the state of a system is given by a density matrix $\rho_{\theta}$ instead of a probability distribution function. Random data will be collected after the measurements on the system and these data could be used to reconstruct the parameter $\theta$ of the system. Given a positive-operator-valued measurement (POVM) denoted by a positive operator $M_x$, where x labels the possible measurement outcomes, the classical probability $p_{\theta}(x)$ in equation \ref{cfi} is given by $p_{\theta}(x)=\text{tr}(\rho_{\theta}(x) M_x)$.
By introducing the symmetric logarithmic derivative $L_{\rho_\theta}$, the Fisher information for $M=1$ in \ref{cfi} can be written as
\beq
  J_\theta=\int dx\frac{1}{\text{tr}(\rho_\theta M_x)}\left(\text{tr}\left(\frac{1}{2}(\rho_\theta L_{\rho_\theta}+L_{\rho_\theta}\rho_\theta) M_x\right)\right)^2\,,
\eeq
where $\frac{\partial\rho_\theta}{\partial \theta}=\frac{1}{2}(\rho_\theta L_{\rho_\theta}+L_{\rho_\theta}\rho_\theta)$ and $L_{\rho_\theta}=2\int_{0}^{\infty} ds e^{-s \rho_\theta} \frac{\partial\rho_\theta}{\partial \theta}e^{-s \rho_\theta}$.
The maximum of all possible measurements gives a measurement independent quantity $\mathcal{F}_\theta $ which is termed as the \textit{quantum Fisher information} (QFI) \cite{braun2018quantum},
\begin{equation}
 \mathcal{F}_\theta \equiv \rm{tr} (\rho_\theta L_{\rho_\theta}^2)=\rm{tr}( L_{\rho_\theta} \frac{\partial\rho_\theta}{\partial \theta})\,.
\end{equation}
The quantum version of the classical Cram\'er-Rao (lower) bound is called the quantum Cram\'er-Rao bound, which reads $\text{Var}(\theta_\text{est}) \geq \frac{1}{M \mathcal{F}_\theta}$ where M is the number of independent identical POVM measurements of the same system prepared in the same state. Those measurements that can saturate the quantum Cram\'er-Rao bound are called optimal distinguishing measurements \cite{hyllus2012fisher,braun2018quantum,yao2014quantum}. The close connection between the QFI and the Bures distance is the following. Bures distance endows the Riemannian geometry onto the space of density operators, $d_B(\rho_1,\rho_2)=\sqrt{2-2A(\rho_1,\rho_2)}$, where $A(\rho_1,\rho_2)=\text{tr}(\rho_1^{1/2}\rho_2\rho_1^{1/2})^{1/2}$ represents the fidelity \cite{yao2014quantum,braunstein1994statistical}. It can be expressed in the QFI through the integration of the infinitesimal distance $d_B(\rho_{\theta},\rho_{\theta+d\theta})=\frac{1}{4} \mathcal{F}(\rho_{\theta})d^2 \theta$ for two density operators $\rho_{\theta}$ and $\rho_{\theta+d\theta}$ \cite{lu2013broadcasting}.

Quantum correlations measure the degree of cooperative responses from two quantum systems. Many such measures exist in the literature, such as the entanglement, quantum discord and quantum mutual information \cite{hill1997entanglement,wootters1998entanglement,jaeger2007quantum,ollivier2001quantum}. The entanglement formation, or concurrence, for a four dimensional density matrix is given by $\mathcal{E(\rho)}=\text{Max}(0,\sqrt{\lambda_1}-\sqrt{\lambda_2}-\sqrt{\lambda_3}-\sqrt{\lambda_4)}$, 
where $\lambda_i$ are the eigenvalues of the Hermitian matrix R in the descending order where R is defined as $R=\rho(\sigma_y\otimes\sigma_y)\rho^*(\sigma_y\otimes\sigma_y)$.
In our case of study, the relevant ones are the ``X" type density matrices. The concurrence then reduces to the following expression,
\beq \mathcal{E(\rho)}=2\ \text{Max}(0,|\rho_{23}|-\sqrt{\rho_{11}\rho_{44}},|\rho_{14}|-\sqrt{\rho_{22}\rho_{33}}). \eeq

Quantum mutual information (QMI) between two subsystems A and B is a direct generalization of classical mutual information and is defined as:
\begin{eqnarray}
\mathcal{I} (\rho^{AB}) = S (\rho^A) + S (\rho^B) - S(\rho^{AB})\ , \label{QMI}
\end{eqnarray}
where $S(\rho) = - \mathrm{tr} \, ( \rho \, \log_2 \rho )$ is the von Neumann entropy and $\rho^{A(B)}$ is the reduced density operator of the subsystem $A$($B$), respectively. Another generalization of the classical mutual information is the classical correlation $\mathcal{C}(\rho)$ and it is defined as the superior of $\mathcal{I}(\rho|\{B_k\})$ over all possible von Neumann measurements where
\begin{eqnarray}
\mathcal{I}(\rho|\{B_k \}):=S(\rho^A)-S(\rho|\{B_k\})\,,
\end{eqnarray}
and $S(\rho|\{B_k\}):=\sum_{k}p_kS(\rho_k)$ is the quantum conditional entropy with respect to the von Neumann measurements ${B_k}$. The post-measurement  conditional density operator $\rho_k$ associated with the measurement result $k$ is the $\rho_k=\frac{1}{p_k}(I\otimes B_k)\rho (I\otimes B_k)$,
where $p_k=tr(I\otimes B_k)\rho (I\otimes B_k)$ and $I$ is the identity operator. The quantum discord $\mathcal{Q}$, which reflects the non-classical correlations between the two subsystems is defined to be the difference between the quantum mutual information and the classical correlation, $\mathcal{Q}(\rho):=\mathcal{I}(\rho)-\mathcal{C}(\rho)$.

\vspace{1cm}
\section{Formalism}
\begin{figure}[ht]
\centering
\includegraphics[width=0.5\textwidth]{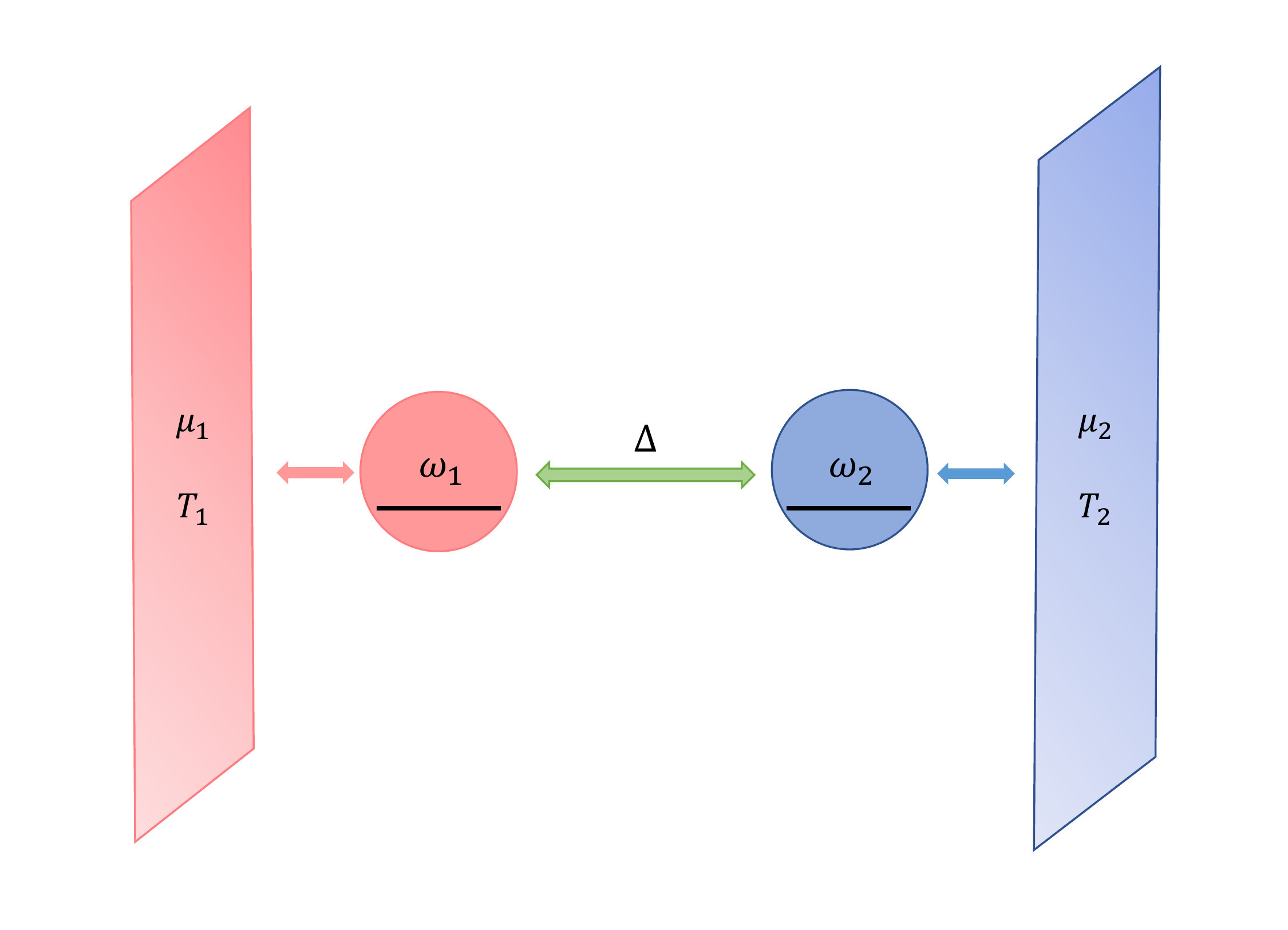}
\caption{Schematic demonstration of the system.}
    \label{theme}
\end{figure}

To address the the effect of nonequilibrium environments, we consider a simplified molecular junction model. The schematic demonstration of the system in shown in Fig.~\ref{theme}. This setup is pertinent to the essential features of molecular junctions, charge and energy transport and spin-chain systems in non-equilibrium environments as mentioned in the previous section. We take the Hamiltonian of the above system to be as follows:
\beq \calh_S=\omega_1 \eta_1^\dagger \eta_1+\omega_2 \eta_2^\dagger \eta_2+\Delta(\eta_1^\dagger \eta_2+ \eta_2^\dagger \eta_1) \\
\eeq
where $\calh_S$ represents the free system Hamiltonian without the environments. For the two fermionic sites (S1, S2) each can either adopt a fermon with certain energy $\omega_i$ or remain an empty state. Fermion(s) on S1 and S2 have a finite rate of hopping between the two sites with strength $\Delta$. The two sites are immersed in two separate large reservoirs (two electrodes) described by the following Hamiltonian,
\beq 
\calh_R=\sum_{k,p}\hbar \omega_k \ (a_{kp}^\dagger a_{kp})+\sum_{q,s}\hbar \omega_q \ (b_{qs}^\dagger b_{qs}).
\eeq

The creation and annihilation operators $\eta_{1,2}^\dg$ and $\eta_{1,2}$ on the site 1(2) follow the standard fermionic statistics. The interaction between the system and the reservoirs is the local process of the subsystem S1 or S2 emitting a fermion to the environment or taking a fermion from the bath it is in contact with. The Hamiltonian can be written as the following form,
\beq \mathcal{H}_{int}=\sum_{k,p}\lambda_k \ (\eta_1^\dagger a_{kp}+ \eta_1 a^\dagger_{kp})+\sum_{q,s} \lambda_q \ (\eta_2^\dagger b_{qs}+ \eta_2 b^\dagger_{qs}), \eeq
where $\lambda$ is the interaction strength between the system and the reservoir and $a_{kp}^{\dg} (b_{kp}^\dg)$ are the creation operators of a particle with momentum $k$, polarization $p$ in the reservoirs, while $a_{kp} (b_{kp})$ are the corresponding annihilation operators. The system Hamiltonian can be diagonalized with the transformation, 
\beq \vec{\zeta}=\begin{pmatrix} \cos(\theta/2) & \sin(\theta/2) \\ -\sin(\theta/2) & \cos(\theta/2) \end{pmatrix} \vec{\eta} \label{newH}\eeq
where cos$\theta=\dfrac{w_2-w_1}{\sqrt{(w_1-w_2)^2+4 \Delta^2}}$. After the diagonalization, \beq \calh_S=\omega'_1 \zeta_1^\dagger \zeta_1+\omega'_2 \zeta_2^\dagger \zeta_2 \eeq with $\omega'_{1,2}=\frac{1}{2} (\omega_1+\omega_2 \pm \sqrt{(\omega_1-\omega_2)^2+4 \Delta^2})$ and $\zeta_{1,2}$ ($\zeta_{1,2}^\dg$) satisfy the anti-commutation relation.

We apply the quantum master equation approach with the Born-Markov approximation and without secular approximation. For details in this approach and the caveat used in our model, see \cite{wang2019nonequilibrium}. We ignore the variance in the reservoirs, i.e. $\til \rho_R(t) \approx \rho_R(0)=\rho_R$ and the quantum master equation (QME) takes the following form,
\beq \dfrac{d \tilde \rho_S(t)}{dt} =-\mathrm{Tr}_R \int_0^t ds \left[\tilde H_{int}(t),[\tilde H_{int}(s),\til \rho_S(t)\otimes \rho_R]\right] \eeq
where $\til{H}_{int}\equiv e^{iH_0t/\hbar}H_{int}e^{-iH_0t/\hbar}$ is the interaction Hamiltonian in the interaction picture, $H_{int}$ is the interaction Hamiltonian in the Schr\"odinger picture and $H_0=H_S+H_R$. We ignore the back reaction from the reservoir to our system and trace out the environmental contributions to the full density matrix to obtain the reduced density matrix of our system.

The secular approximation used in the Lindblad master equations fails to capture the population and coherence couplings which is crucial for nonequilibrium studies \cite{wang2019nonequilibrium,zhang2014curl,wang2019steady}. Moreover, secular approximation requires extra time scale hierarchy that is not always applicable, in particularly when nonequilibrium effects are prominent \cite{wang2019nonequilibrium,li2015steady}. Generally speaking, the secular approximation is not needed when the intra-system coupling (or the hopping rate) is much less than the levels of the system, $\Delta\ll \omega_i$. The discussion of the applicability of different master equations and the secular approximation can be found in Refs. \cite{hofer2017markovian,gonzalez2017testing,carmichael1973master}. In this study, we apply Markovian approximation without secular approximation (Bloch-Redfield equation) and examine the nonequilibrium impacts on the system QFI, the quantum correlations characterized by coherence, linear entropy, mutual information and quantum discord, and the thermodynamic cost characterized by the entropy production. The explicit form of the equation is given in the Appendix.~\ref{append.a}.

\section{QFI of X-type bipartite systems and EPR}
For the system we consider, we are interested in the long time steady state behavior, namely the nonequilibrium steady state solution (NESS). The solution for the density matrix is obtained by solving the quantum master equation Eq.~\ref{mst} given in the appendix,
\beq \dot \rho_S(t)=i[\rho_S,H_S]-D_0[\rho]-D_s[\rho]=0\,. \eeq
The density matrix of the NESS has two uncoupled parts, of which one is the nonzero entries given in Eq.~\ref{rhomatrix}, and the rest are off-diagonal coherence components which are uncoupled with the population terms thus vanish in the long time limit. Therefore, we only need to consider the "X" form solution in the energy representation,
\bea  \rho_S(\infty)=\begin{pmatrix}\rho_{11} & 0 & 0 & 0 \\ 0& \rho_{22} & \rho_{23} & 0 \\ 0&\rho_{32}& \rho_{33}&0\\0&0&0&\rho_{44} \end{pmatrix}\,,\label{rhomatrix}\eea
where $\rho_S(\infty)$ denotes the NESS solution and will be shorthanded as ``$\rho$" from now on. One feature of the Bloch-Redfield equation is the non-vanishing off-diagonal elements $\rho_{23}$ and $\rho_{32}$, the coherence terms in the energy basis which are omitted by the Lindblad formalism. Those terms are unimportant in the equilibrium regimes but are crucial in the description of nonequilibrium physics. 

After performing the spectral decomposition $\rho=\sum_{i}^{N}p_i|\psi_i\rangle \langle\psi_i|$ (where $N$ denotes the number of nonzero $p_i$), the QFI for a general quantum state described by the density matrix $\rho(\theta)$ is given by \cite{knysh2011scaling}
\begin{widetext}
\begin{equation}\label{fishform}
\mathcal{F}_{\theta}=\sum_{i=1}^{N}\frac{(\partial_{\theta}p_{i})^{2}}{p_{i}}+
4\sum_{i=1}^{N}p_{i}\langle\partial_{\theta}\psi_{i}|\partial_{\theta}\psi_{i}
\rangle-\sum_{i,j=1}^{N}\frac{8p_{i}p_{j}}{p_{i}+p_{j}}|\langle\psi_{i}|
\partial_{\theta}\psi_{j}\rangle|^{2}.
\end{equation}
\end{widetext}
The coefficients of the spectrum decomposition in \ref{fishform} can be given in terms of the elements of the reduced density matrix and is given as follows,
\begin{widetext}
\begin{eqnarray}\label{spec}
p_{1} & = & \rho_{11},\quad |\psi_{1}\rangle=|g\rangle,\\
p_{2} & = & (\rho_{22}+\rho_{33})/2+\sqrt{(\rho_{22}-\rho_{33})^{2}/4+|\rho_{23}|^{2}},\quad |\psi_{2}\rangle=\cos\frac{\alpha}{2}e^{i\phi}|e\rangle+\sin\frac{\alpha}{2}|f\rangle,\\
p_{3} & = & (\rho_{22}+\rho_{33})/2-\sqrt{(\rho_{22}-\rho_{33})^{2}/4+|\rho_{23}|^{2}}, \quad |\psi_{3}\rangle=\sin\frac{\alpha}{2}e^{i\phi}|e\rangle-\cos\frac{\alpha}{2}|f\rangle,\\
p_4 & = & \rho_{44}, \quad |\psi_{4}\rangle=|f\rangle,
\end{eqnarray}
\end{widetext}
where
\begin{equation}
\alpha=\arctan[2|\rho_{23}|/(\rho_{22}-\rho_{33})], \quad \phi=\arg(\rho_{23}).
\end{equation}

For the bipartite quantum system with the spectral decomposition given by Eq.~(\ref{spec}), the QFI according to Eq.~(\ref{fishform}) can be written as follows,
\begin{equation}\label{QFI}
\mathcal{F}_{\theta}=\sum_{i=1}^{4}\frac{(\partial_{\theta}p_{i})^{2}}{p_{i}}
+\frac{(p_2-p_3)^2}{p_2+p_3}\left[(\partial_{\theta}\alpha)^{2}+(\partial_{\theta}\phi)^{2}\sin^2\alpha\right]\,.
\end{equation}
We denote the expression of the QFI as $\mathcal{F}_\theta= \mathcal{F}_{\theta}^{E} + \mathcal{F}_{\theta}^{N}$, where $\mathcal{F}_{\theta}^{E}=\sum_{i=1}^{4}\frac{(\partial_{\theta}p_{i})^{2}}{p_{i}}$ and $\mathcal{F}_{\theta}^{N}=\frac{(p_2-p_3)^2}{p_2+p_3}\left[(\partial_{\theta}\alpha)^{2}+(\partial_{\theta}\phi)^{2}\sin^2\alpha\right]$, for the reason to be discussed in the following text. 

When the two reservoirs are identical, i.e. $T_1=T_2=T$ and $\mu_1=\mu_2=\mu$, the system relaxes to the equilibrium with the reservoirs, and the reduced density matrix in energy basis is given as follows, 
\beq  \rho_S(\infty)=\begin{pmatrix}\dfrac{e^{2 \hbar \beta (\omega-\mu)}}{Z} & 0 & 0 & 0 \\ 0& \dfrac{e^{\hbar \beta \omega'_2-\mu}}{Z} & 0 & 0 \\ 0 & 0 & \dfrac{e^{\hbar \beta \omega'_1-\mu}}{Z} & 0 \\ 0 & 0 & 0 & \dfrac{1}{Z} \end{pmatrix}\, ,\eeq
where $\beta=\frac{1}{T}$ is the reciprocal of the temperature and $Z=\left( e^{2 \hbar \beta (\omega-\mu)}+e^{\hbar \beta (\omega'_2-\mu)}+e^{\hbar \beta (\omega'_1-\mu)}+1\right)$ is a normalization factor. The coherence terms of the reduced density matrix for the system vanish for the equilibrium case. For the symmetric junction case, $\omega_1=\omega_2=\omega$, the QFI can be expressed as follows in the weak tunneling limit $\Delta\ll \omega$,
\beq \mathcal{F}_{\Delta}\approx \frac{\beta^2}{Z}(e^{\beta(\omega+\Delta-\mu)}+e^{\beta(\omega-\Delta-\mu)})=\frac{2 \beta^2}{Z}e^{\beta(\omega-\mu)} \cosh\Delta \, ,
\eeq  
and $Z\approx (1+e^{\hbar \beta(\omega-\mu)})^2$. For equilibrium states with equal temperature and chemical potential for the two reservoirs, only the first term $\mathcal{F}_{\theta}^{E}$ in Eq.~\ref{QFI} survives. The second term $\mathcal{F}_{\Delta}^{N}=\frac{(p_2-p_3)^2}{p_2+p_3}\left[(\partial_{\theta}\alpha)^{2}+(\partial_{\theta}\phi)^{2}\sin^2\alpha\right]$ vanishes in the equilibrium condition as the steady-state coherence $\rho_{23}=0$ and $\alpha=0$. We notice that the first term in Eq.~(\ref{QFI}) is formally equivalent to the classical Fisher information if we identify the set of eigenvalues as the probability distribution. The remaining parts correspond to the contribution from the appearance of the quantum coherence, which is absent in the equilibrium scenario.

On the other hand, for the two environments not in equilibrium, i.e. $T_1\ne T_2$ or $\mu_1\ne \mu_2$, the nonzero terms of the reduced density matrix up to the first order in the expansion of $\Gamma/\Delta$ are given as below,
\beq \bsplit  &\rho_{11}= (1-n_{1p}) (1-n_{2p})+\mathcal{O}(g^2)\,,\\
&\rho_{22}=( n_{1p} - n_{1p} n_{2p})+\mathcal{O}(g^2)\,,\\
&\rho_{33}=( n_{2p} - n_{1p} n_{2p})+\mathcal{O}(g^2)\,,\\
&\rho_{44}=n_{1p} n_{2p}+\mathcal{O}(g^2)\,,\\
&\rho_{23}=-i(n_{1m} + n_{2m}) g/2+\mathcal{O}(g^2)\,,\\
&\rho_{32}=i(n_{1m} + n_{2m}) g/2+\mathcal{O}(g^2)\,, \end{split} \label{solution}\eeq
where $n_{i,p/m}=(n(\omega'_i,T_1,\mu_1)\pm n(\omega'_i,T_2,\mu_2))/2$, $g=\frac{\Gamma}{\Delta}$ and $n(\omega'_i,T_j,\mu_j)$ follows Fermi-Dirac distribution. The distinct trait of this solution is the nonvanishing coherence terms $\rho_{23}$ and its complex conjugate $\rho_{32}$. It is worth noticing that both of the two terms in Eq.~\ref{QFI} contribute to $\mathcal{F}_{\Delta}$ in the nonequilibrium conditions in contrast to that in the equilibrium scenario. This contribution from the nonequilibrium coherence is non-negative and thus contributes positively to the QFI.

\subsection*{Nonequilibrium current and thermodynamic cost}
This enhanced coherence from the nonequilibrium conditions can appear in many places, such as the entanglement and quantum correlations of the two subsystems and QFI's, but this enhanced quantum effect comes with the thermodynamic cost characterized by the entropy production. The entropy production rate (EPR) that comes with the enhanced quantum effect is a measure of nonequilibriumness. Experimentally, it represents the degree of disorder and randomness induced in the system, which may be important for precise measurement. The relationship between the EPR and the QFI is what we focus on in this paper.

The nonequilibrium process often lead to the quantum flux of energies between the two subsystems and the two reservoirs. The energy flux reaches a constant value when the NESS is reached. We consider the case with fermionic reservoirs. In the NESS, the constant flux is in the form of particle exchange between the reservoir and the system. The particle flows from the reservoir with a high temperature or chemical potential to the bath with a low temperature or chemical potential. The particle flux is characterized by the following expression,
\begin{equation}
    \tr(\dot\rho_S \mathcal{N}_S) = \tr\left(\left(i[\rho_S,H_S]+\sum_{i=1,2}D_i[\rho_S]\right)\mathcal{N}_S\right)= \sum_{l=1}^2 I_l\,,
\end{equation}
where $I_l$ with $l=1,2$ are the particle currents from the reservoir 1 or 2 to the system, and $\mathcal{N}_S=\zeta_1^\dagger \zeta_1+\zeta_2^\dagger \zeta_2$ is the number operator which commutes with the system Hamiltonian $\calh_S$, i.e. $[\mathcal{N}_S,H_S]=0$. The commutator term is the unitary evolution of the state operator under the free system Hamiltonian and does not contribute in the expression of the current. It can be easily shown that the corresponding term vanishes after taking the trace and applying the cyclic identity of matrix tracing due to the commutation relation of the number operator with the system Hamiltonian. According to the master equation, we have
\begin{equation}
    I_l = \tr(D_l[\rho_S]\mathcal{N}_S)\, ,\label{parcurnt}
\end{equation}
where the steady state dissipators $D_l[\rho_S]$ are given by the Bloch-Redfield equation
\beq
\dot\rho_S(t)=i[\rho_S,H_S]+\sum_{i=1,2}D_i[\rho_S]\,.
\eeq
Here, $D_i[\rho_S]$ is the dissipator in contact with the $i^{\mathrm{th}}$ reservoir which represents the particle current from that reservoir. In the Lindblad form, only diagonal (equal frequency) operators show up as the result of the secular approximation. In the Redfield form, the fast processes (crossing terms) which represent the coherence contribution to the dynamics are considered and not thrown away. It has the same interpretation as that of the Lindbladian superoperators with additional coherence contributions. The explicit expression is given in the Appendix \ref{append.a}. Similarly, we can define the energy current $J_l$:
\begin{equation}
    \tr(\dot\rho_S \calh_S) = \sum_{l=1}^2 J_l\,,
\end{equation}
where
\begin{equation}
    J_l = \tr(D_l[\rho_S]\calh_S)\,.
\end{equation}
In the NESS, the current from the first reservoir to the system is equal to that from the system to the second reservoir, and it can be shown that $J_1+J_2=0$ and $I_1+I_2=0$. Since the two thermal reservoirs are only connected by a microscopic quantum system, the energy transfer process is assumed to be slow and the bath is assumed to be classical with short thermalization time. We assume the validity of the semi-classical description of the entropy production, and the EPR of the environment is approximately given as follows,
\begin{align}
    \dot S&=-\frac{J_1-\mu_1I_1}{T_1}-\frac{J_2-\mu_2I_2}{T_2}\nonumber \\
    &=-J_1\left(\frac{1}{T_1}-\frac{1}{T_2}\right)+I_1\left(\frac{\mu_1}{T_1}-\frac{\mu_2}{T_2}\right)\,.
    \label{epr}
\end{align}
In order for the reservoirs to remain constant in the NESS, external work needs to be implemented to extract the entropy away from the system. Therefore, the entropy increase of the external systems is at least at the rate given by Eq.~\ref{epr}. We also remind that the semi-classical approximation of the entropy production can have negativity problem in certain cases using the local master equations, e.g. when the inter-site coupling is strong. This naively appears to violate the second law of thermodynamics \cite{levy2014local}. This issue arise due to that the coarse-grained definition of EPR is a semi-classical approximation and cannot be extended into the whole parameter space. In the appendix \ref{positivity}, we provide a proof that the definition works well in the parameter regimes we are interested. We also remark that in the study of quantum thermodynamics, one of the most widely-used definitions of EPR is expressed through the relative entropy, which does not have the problem of negativity in the thermodynamic limit \cite{landi2020irreversible}. However, that definition needs the data about the environment which is not applicable using the master equation approach.

\section{Weak tunneling scenario}
\subsection*{The effect of nonequilibrium current and thermodynamic cost on QFI}

\begin{figure}[htb]
\centering 
\includegraphics[width=0.38 \textwidth]{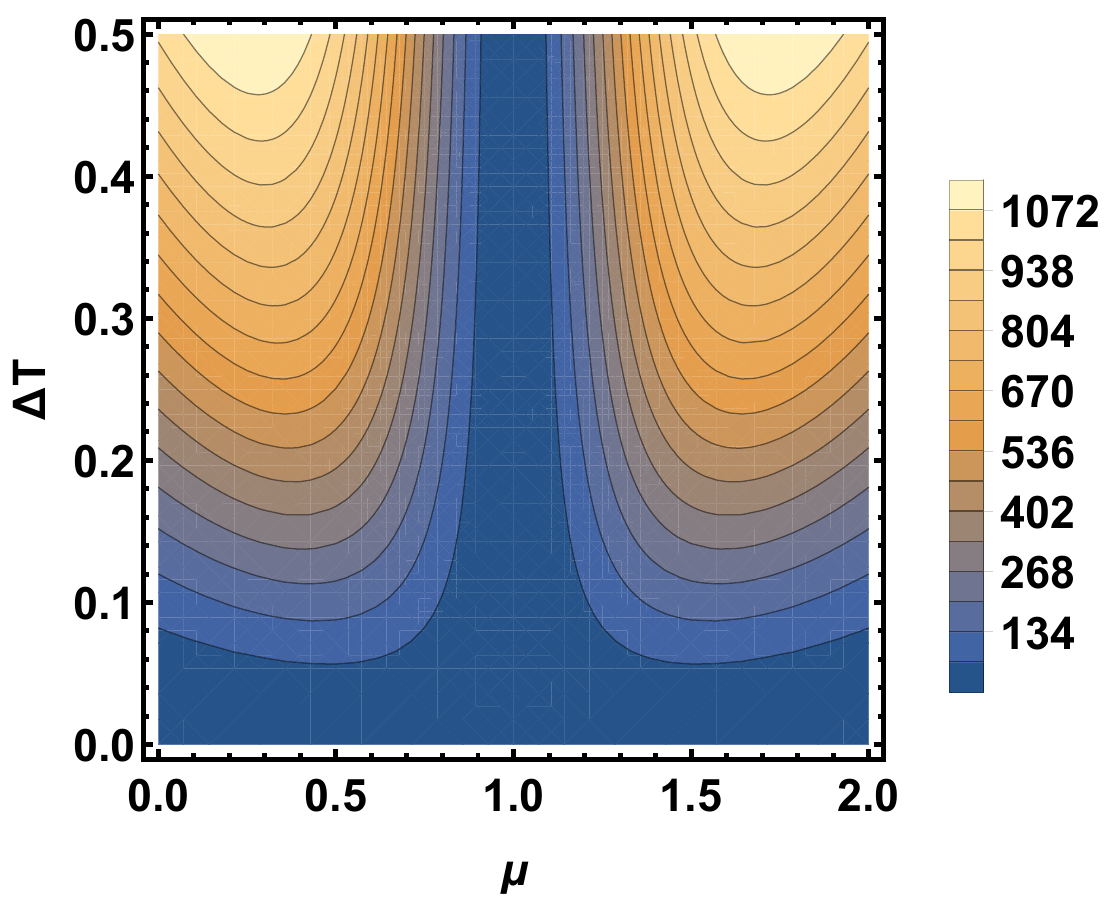}
\quad
\includegraphics[width=0.39 \textwidth]{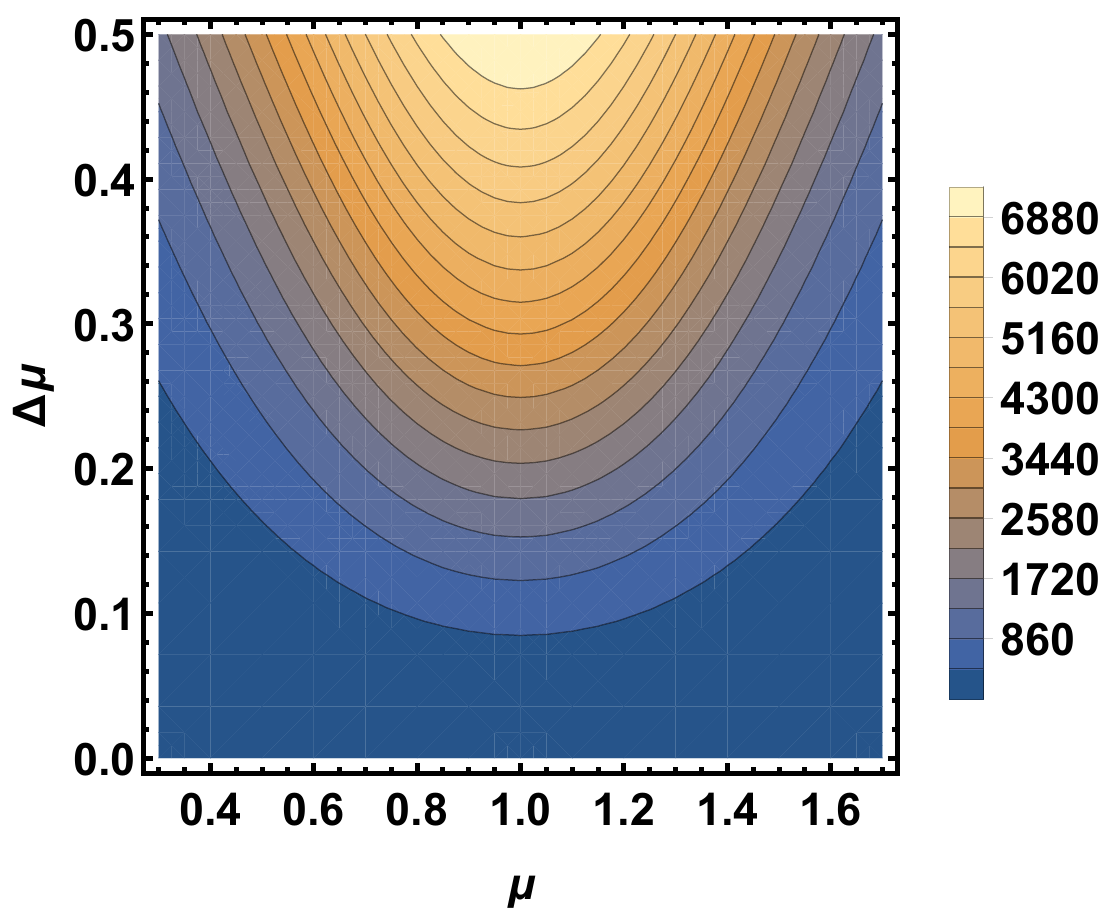}
\caption{QFI in nonequilibrium environments. (a) QFI vs $\Delta T$ at different chemical potentials. $T_1=0.2$, $T_2=T_1+\Delta T$ and $\Delta=0.005$. The QFI is enhanced when the chemical potential is away from system frequencies. (b) QFI vs $\Delta \mu$ at different chemical potentials. $T_1=T_2=0.2$, $\mu_{1,2}=1\pm\Delta \mu$ and $\Delta=0.005$.
For all, $\Gamma_1=\Gamma_2=0.002$, $\omega_1=\omega_2=1$.}
\label{fig.2}
\end{figure}

\begin{figure}[htb]
\centering 
\includegraphics[width=0.38 \textwidth]{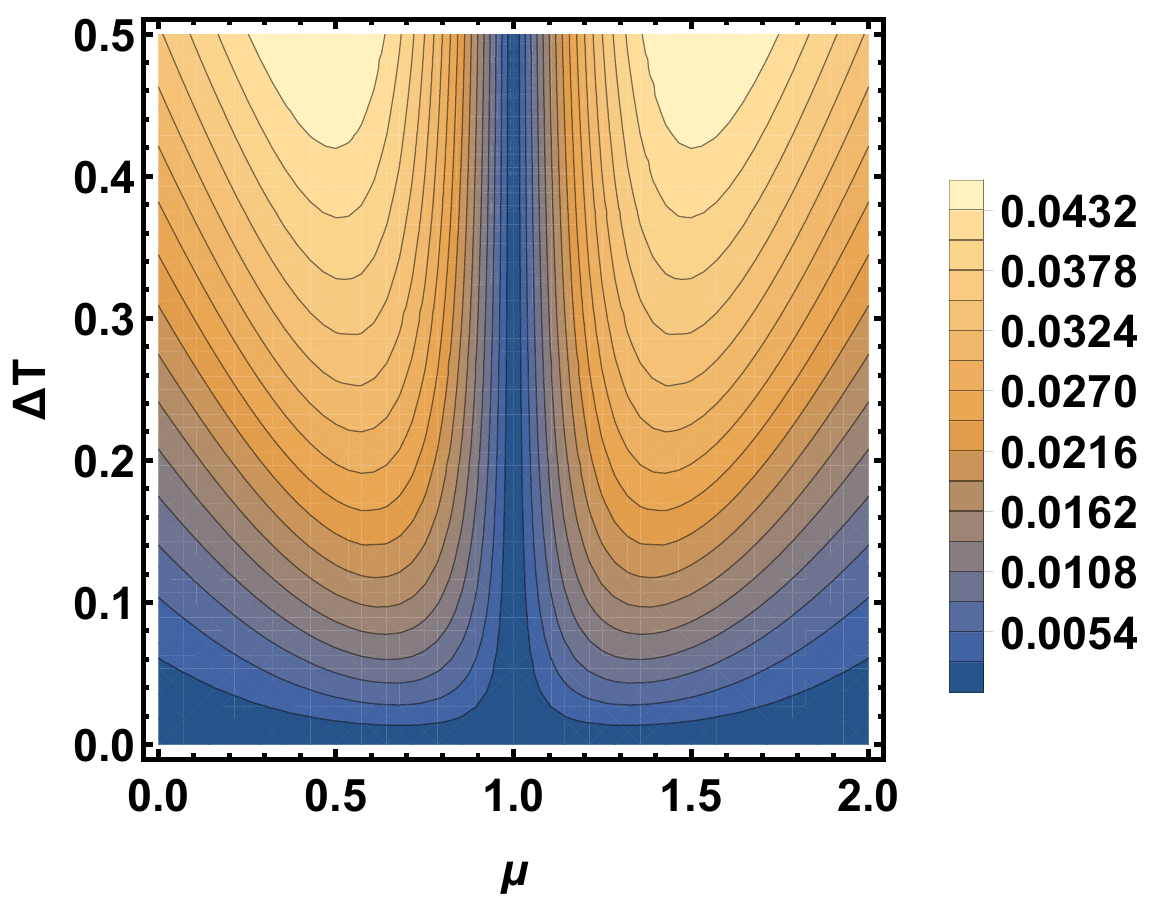}
\caption{Coherence in energy basis. Coherence $|\rho_{23}|$ vs $\Delta T$ at different chemical potentials at the same condition as Fig.~\ref{fig.2} (a).}
\label{fig.2.1}
\end{figure}

In the bipartite quantum system, one of the most important parameters to be finely tuned in experiments is the tunneling rate $\Delta$. Many different methods have been developed to measure this parameter. For example in the model of resonance energy transfer, this parameter can be finely controlled by adjusting the distance between the two parties \cite{jang2004multichromophoric,clapp2006forster}. For a specific quantum measurement, an optimal precision can be achieved when the bias between the two ends (reservoirs) is large, which requires the energy consumption for the external system to maintain the bias. The equilibrium scenario in which the two reservoirs are kept to the same temperature and chemical potential, has been well studied. In general, the two subsystems can be immersed in two different environments and the bias between the two environments will lead to a gradient probability flow from the high temperature or high chemical potential bath to the colder environment or with lower chemical potential. Meanwhile, this energy/probability flow was shown to have a close relation to the quantum coherence and nonlocality in a parpitite system \cite{wang2019steady, wang2019nonequilibrium, yao2014quantum, zhang2014curl}. In the following, we show that the QFI is significantly enhanced in the nonequilibrium environments and that the system dissipates extra entropy in order to maintain the high QFI.

\begin{figure}[htb!]
\centering 
\includegraphics[width=0.31 \textwidth]{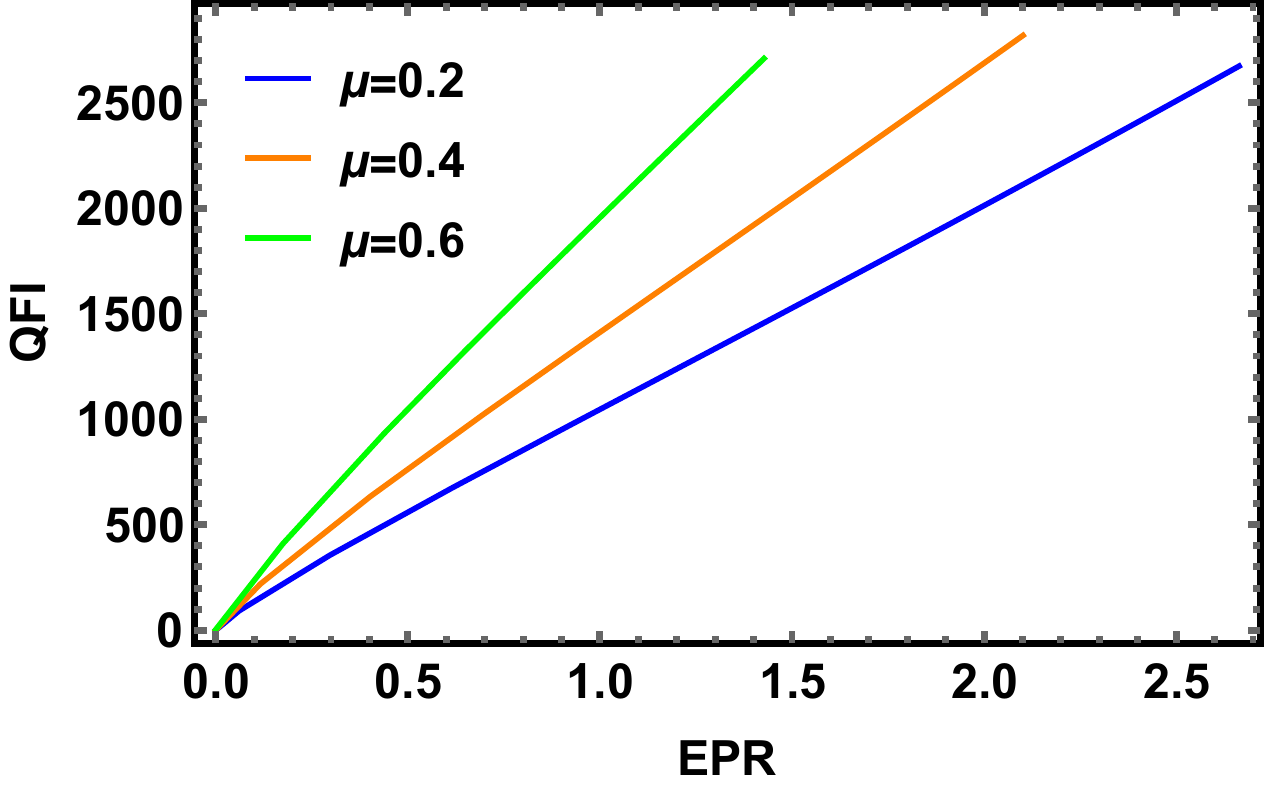}
\quad
\includegraphics[width=0.32 \textwidth]{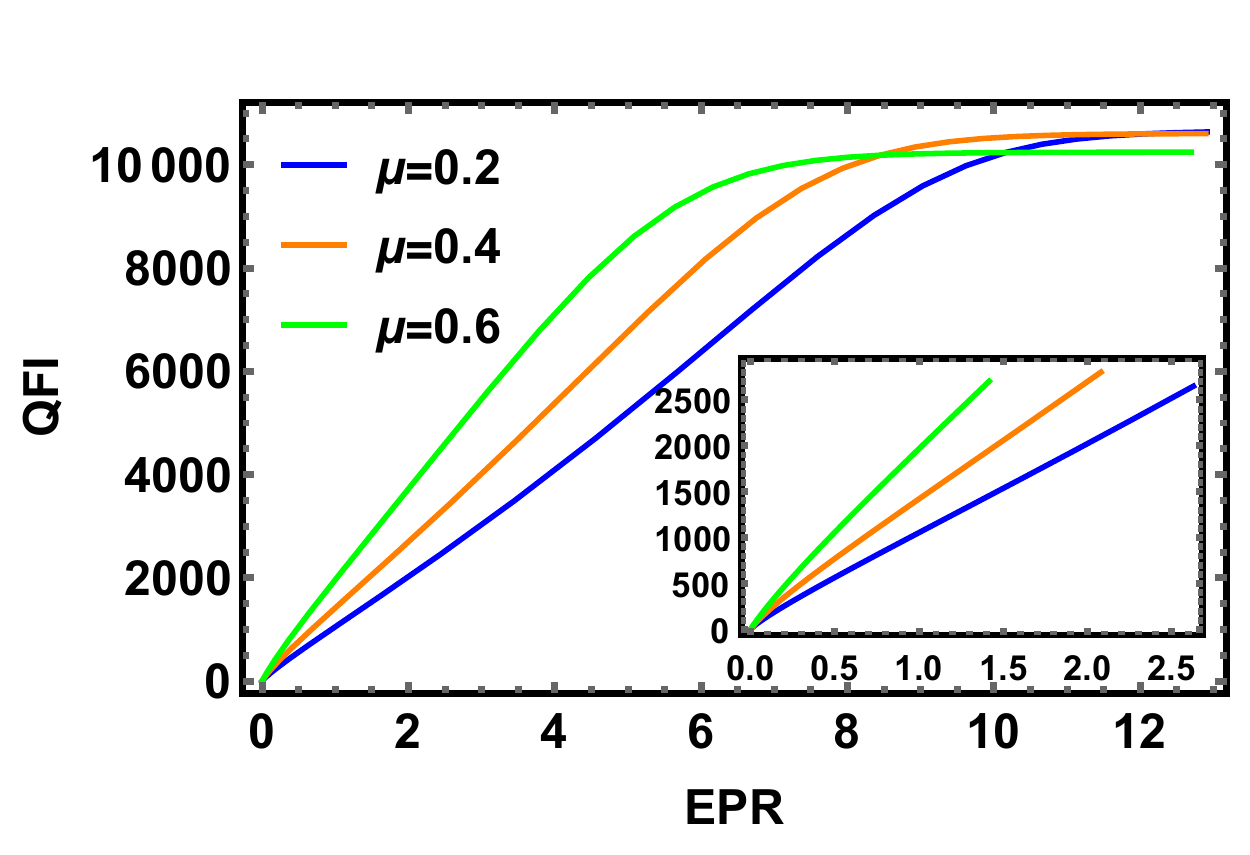}
\quad
\includegraphics[width=0.32 \textwidth]{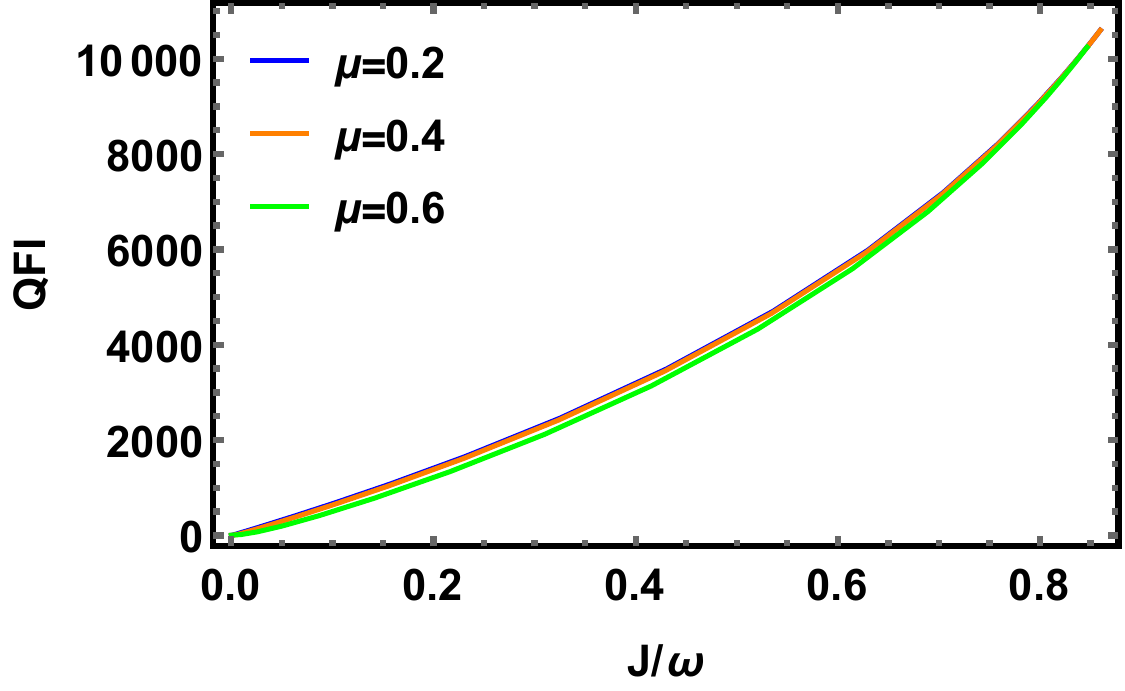}
\caption{QFI and EPR. (a) QFI vs the EPR through enlarging the temperature bias of the two reservoirs at different chemical potentials. $T_1=0.1$ and $T_2=T_1+\Delta T$. (b) QFI vs EPR via unbalancing the chemical potentials of the two reservoirs. $T_1=T_2=0.1$, $\mu_{1}=\mu$ and $\mu_2=\mu_1+\Delta \mu$. (c) QFI vs energy current $J$ via unbalancing the chemical potentials of the two reservoirs. $T_1=T_2=0.1$, $\mu_{1}=\mu$ and $\mu_2=\mu_1+\Delta \mu$. For all, $\Gamma_1=\Gamma_2=0.002$, $\omega_1=\omega_2=1$, and $\Delta=0.005$.}
\label{fig.4}
\end{figure}

In the nonequilibrium regime with weak tunneling, the temperature bias and the chemical potential bias, which are treated as measures of the degree of the nonequilibrium, appear to be strongly positively correlated with the QFI. This can be used as a resource for quantum metrology. As is shown in Fig.~\ref{fig.2} (a), the QFI at a small tunneling rate $\Delta=0.005$ is plotted against the temperature bias $\Delta T$. In the regime away from the resonance chemical potential $\mu=1$, the QFI rises sharply with the increase of the temperature bias compared to its equilibrium values which lie at the bottom of the figures. The QFI reaches a finite asymptotic value as $\Delta T \rightarrow \infty$. When modulating the chemical potential bias, we find the universal nonequilibrium enhancement in the parameter regimes we study without the in-resonance anomaly. The QFI against the chemical potential bias is shown in  Fig.~\ref{fig.2} (b). The nonequilibrium enhancement is easily noticed by comparing the QFI's at low $\Delta \mu$'s (lower part of the figure) with those at high $\Delta \mu$'s (upper part of the figure). On the other hand, it should be noted that the nonequlibrium enhancement of the quantum metrology is only applicable when the tunneling rate of the quantum system is weak. For strongly coupled quantum systems, i.e. those with large tunneling rates, the result is dramatically different [see Fig.~\ref{fig.7}]. 

The enhanced QFI is often viewed as due to the entanglement of the system. However, in an nonequilibrium system where the system is often at a mixed state the entanglement vanishes easily. The quantum properties of the system can be measured by other parameters such as the coherence, quantum discord, mutual information or linear entropy. The QFI is enhanced when the mixing of the quantum states is stronger [see Fig.~\ref{fig.2.1}].

The enhancement of the QFI does not come for free. To sustain the system in the NESS, external energy has to be provided to extract the thermodynamic dissipation cost or the entropy produced in the nonequilibrium process. Classically, the temperature or chemical potential gradient appears hand in hand with the energy current and the production of entropy. Quantum mechanically, apart from the energy current the nonequilibrium effect is noted by the enhanced quantum nonlocality and the induced entropy production in the total system. 

In Fig.~\ref{fig.4} we plot the QFI with the EPR of the system. In the NESS, the system remains unchanged in time and the EPR is constant. As can be noticed, the QFI grows monotonically with the increase of the EPR. This suggests that the efficacy of nonequilibrium enhancement of quantum metrology becomes more prominent as more energy is consumed and larger EPR is produced in the process. Tuning the nob of the chemical potentials can pump in the EPR or enhance the QFI much more significantly compared to modulating the temperatures of the reservoirs. No matter which ways we choose to implement the nonequilibrium condition, the same magnification of the QFI is obtained for a given EPR. Since this is an irreversible process, external work is required to keep the reservoirs and the system at their initial states. This suggests that the nonequilibrium entropy production is required for the enhancement of the quantum metrology. In other words, thermodynamic cost or dissipation can be used to increase the precision of estimations in a quantum system.

\subsection*{The effect of nonequilibrium current and thermodynamic cost on quantum correlations}

The QFI often presents itself in tune with the rise and fall of quantum correlations \cite{frowis2019does,kim2018characterizing}. In an open system, the quantum correlations are usually characterized by discord and mutual information, among others. The entanglement as a measure of correlation often vanishes accidentally for mixed states even when the systems are still correlated nonclassically. This phenomenon has been carefully discussed in many previous studies \cite{wang2019nonequilibrium,yu2009sudden}, thus, we will not elaborate more in this paper.

\begin{figure}[htb]
\centering 
\includegraphics[width=0.39 \textwidth]{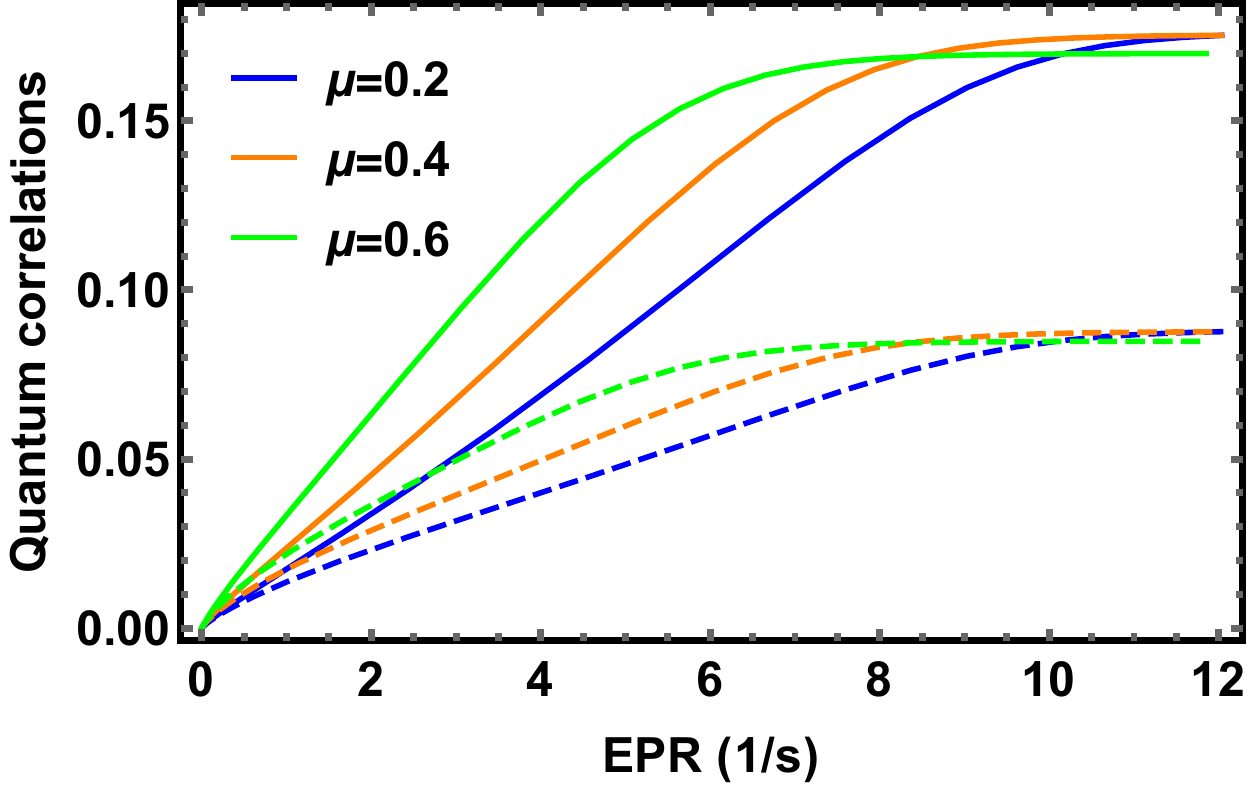}
\quad
\includegraphics[width=0.39 \textwidth]{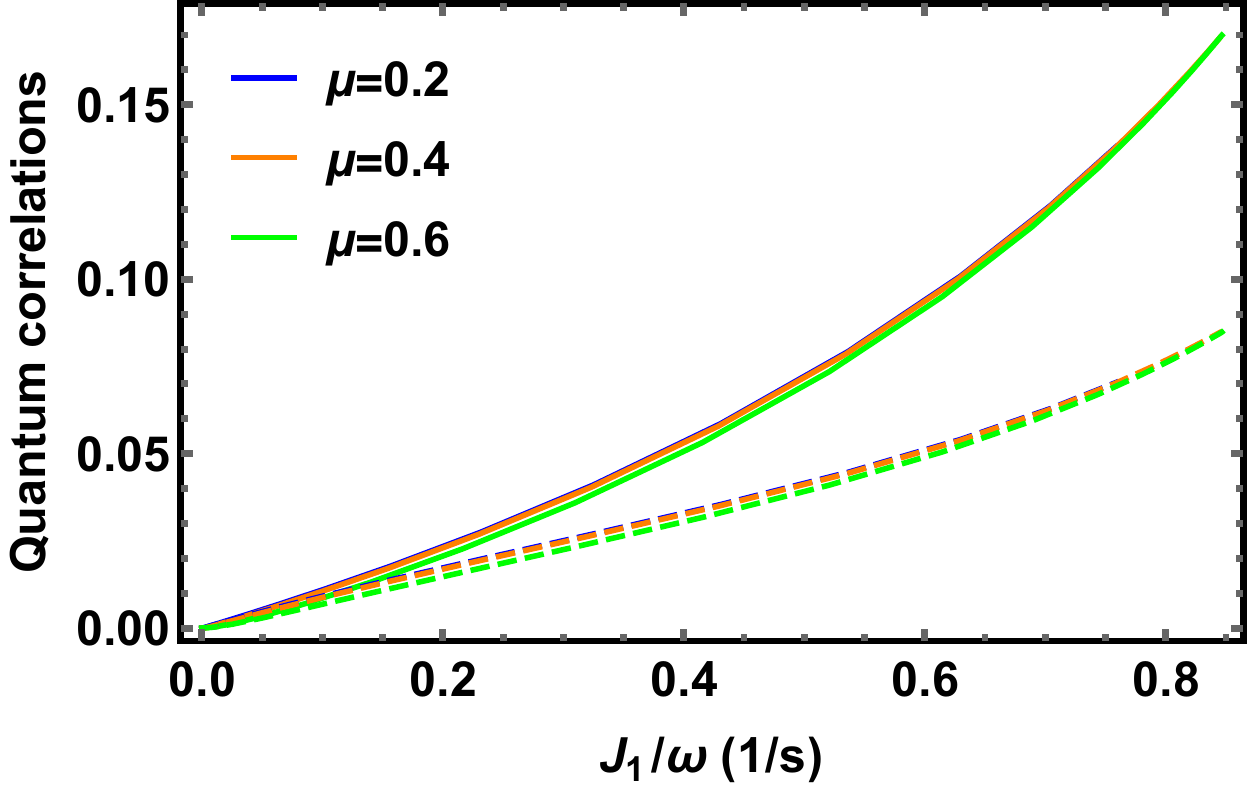}
\caption{(a) Quantum mutual information (solid lines) and quantum discord (dashed lines) vs EPR. (b) Quantum mutual information (solid lines) and quantum discord (dashed lines) vs energy current $J_1$. $T_1=T_2=0.1$, $\mu_1=\mu+\Delta \mu$, $\mu_2=\mu$ and $\Delta=0.005$.}
\label{fig.5}
\end{figure}

In order to obtain strengthened quantum correlations between the two parties in the quantum system, the thermodynamic cost is unavoidable. In the NESS, for the two subsystems placed between the two reservoirs, the strength or the magnitude of the quantum correlations between them shows a simple monotonic relationship with the thermodynamic cost such as the EPR or the energy current. As shown in Fig.~\ref{fig.5} (a, b), the quantum mutual information (QMI) and quantum discord (QD) steadily grow with the increasing EPR and energy current $J_1$. Due to the finite tunneling rate between the two parties in the system, the energy current that flows from one reservoir to the other through the junction saturates at a finite value. The right end of Fig.~\ref{fig.5} (b) represents the asymptotic values for the energy currents, beyond which the system can not be further boosted by only adjusting the chemical potential biases. The quantum correlations stop growing approximately at the EPR where the corresponding energy current reaches the plateau. This shows that in the nonequilibrium open system, the quantum correlations inside the system can be enhanced at the cost of the increased thermodynamic costs characterised through the EPR or the energy current.

\begin{figure}[htb!]
\centering 
\includegraphics[width=0.4 \textwidth]{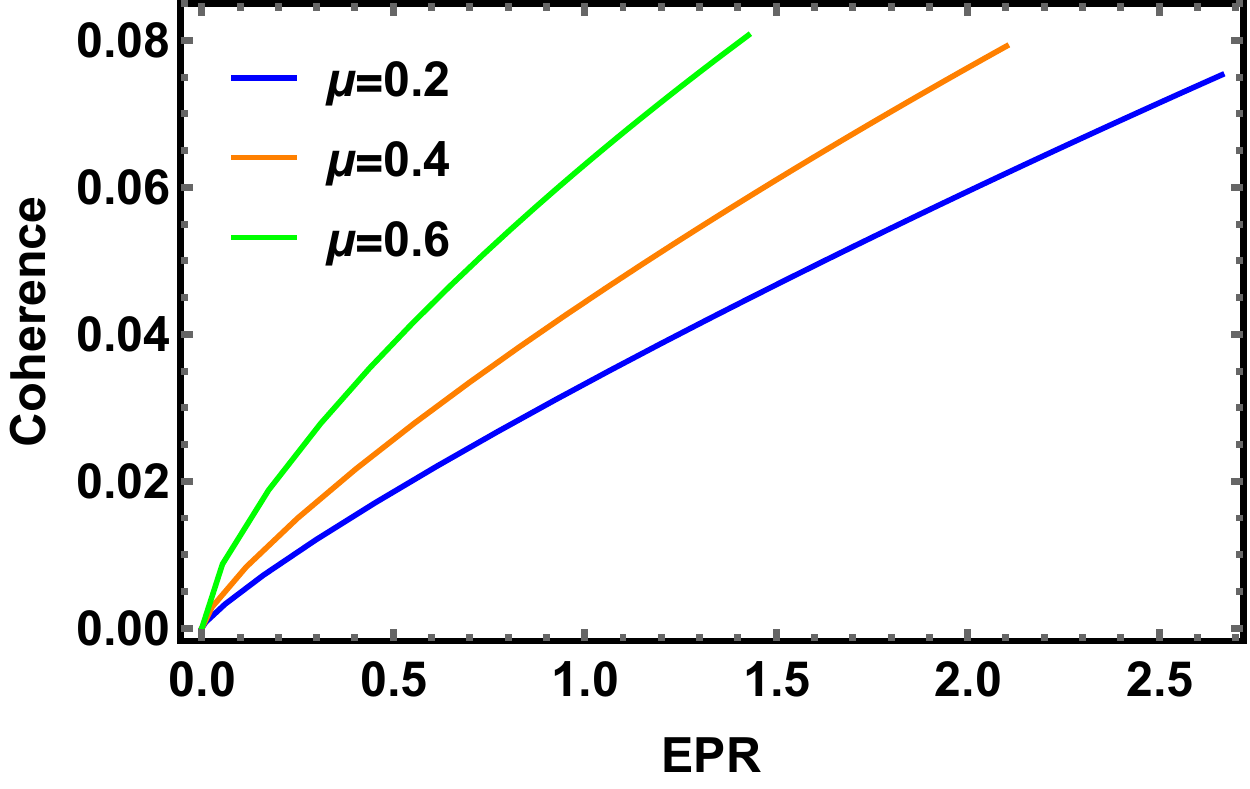}
\quad \quad
\includegraphics[width=0.39 \textwidth]{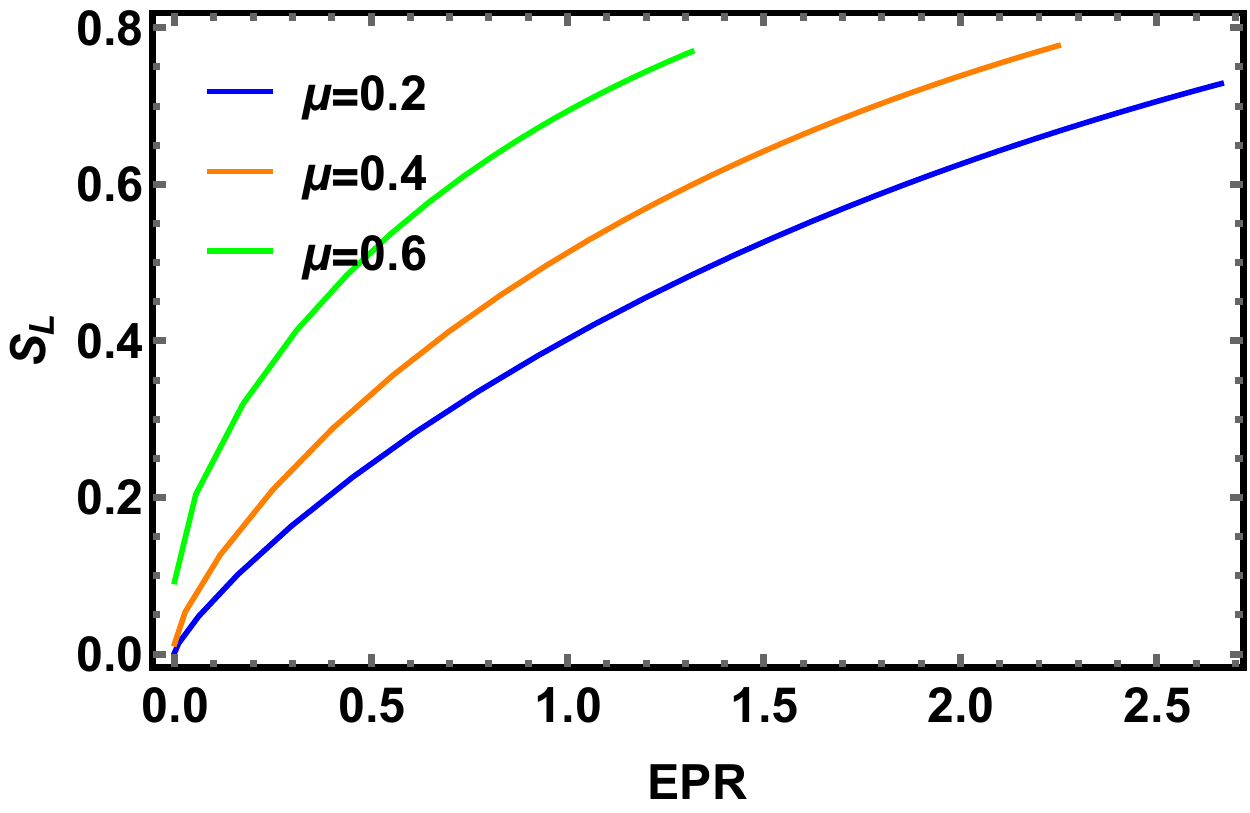}
\caption{(a) Coherence vs EPR enlarging the temperature bias of the two reservoirs at different chemical potentials. (b) Linear entropy $S_L$ vs EPR enlarging the temperature bias of the two reservoirs at different chemical potentials. $T_1=0.1$, $T_2=T_1+\Delta T$ and $\Delta=0.005$.}
\label{fig.6}
\end{figure}

Similarly, the measures of the mixture of the quantum states such as the coherence $|\rho_{23}|$ and the linear entropy $S_L=(4/3) (1-Tr(\hat{\rho_S}^2))$ also increase linearly with the EPR [see Fig.~\ref{fig.6}]. This suggests that the nonequilibrium entropy production is required for the enhancement of quantum correlations and the improvement of the quantum metrology. In other words, the thermodynamic cost or dissipation can be used as a tool to increase the quantum natures of the system.

\section{Strong tunneling scenario}
For quantum systems with strong tunnelings, the impact of nonequilibrium environments on the quantum correlations can be qualitatively different from that of low tunneling rates \cite{wang2019nonequilibrium,wang2018coherence}. We expect the disparity appears in the quantum metrology as well. Regarding the different approaches of master equations, both the local and the global master equations give approximately the same results in this scenario as the tunneling rate falls into the range of $\Gamma\ll \Delta\ll \omega_i$. In our numerical computation, we do not assume the secular approximation. In this section, we discuss the relationship between the thermodynamic cost, the QFI and correlation measures.

\begin{figure}[htb!]
\centering 
\includegraphics[width=0.38 \textwidth]{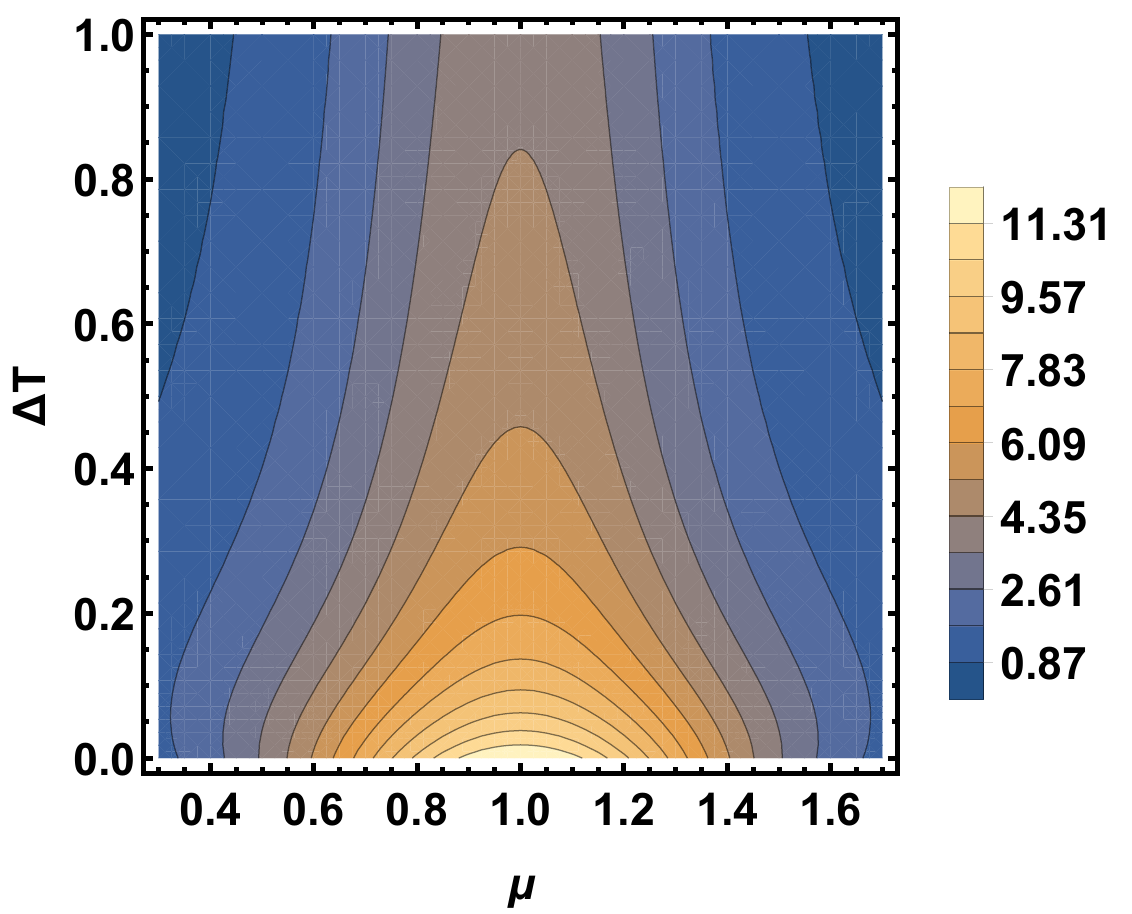}
\includegraphics[width=0.38 \textwidth]{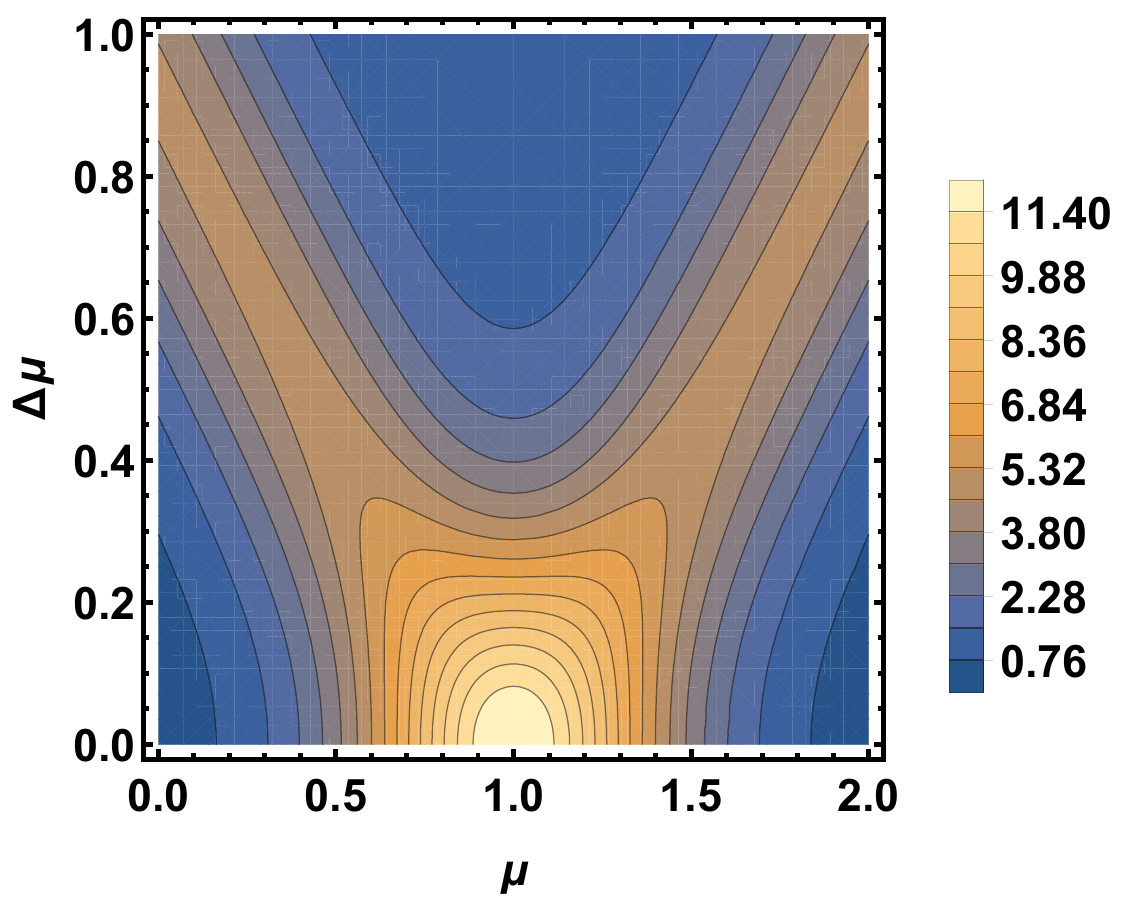}
\caption{QFI in nonequilibrium environments for strong tunneling scenario. (a) $T_1=0.2$, $T_2=T_1+\Delta T$. The nonequilibrium condition suppresses the QFI at large tunnelings. (b) $T_1=T_2=0.2$, $\mu_{1,2}=\mu\pm\Delta\mu$. For both, $\Delta=0.05$, $\Gamma_1=\Gamma_2=0.002$ and $\omega_1=\omega_2=1$.}
\label{fig.7}
\end{figure}

\begin{figure}[htb!]
\centering 
\includegraphics[width=0.314 \textwidth]{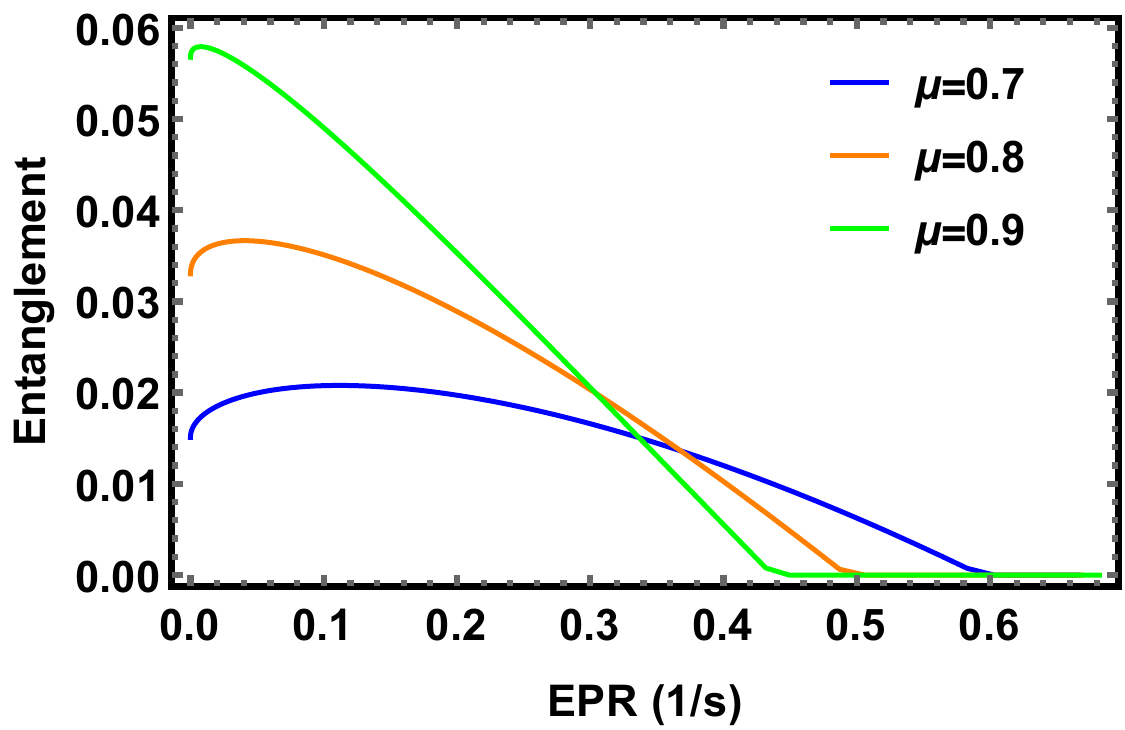}
\includegraphics[width=0.33 \textwidth]{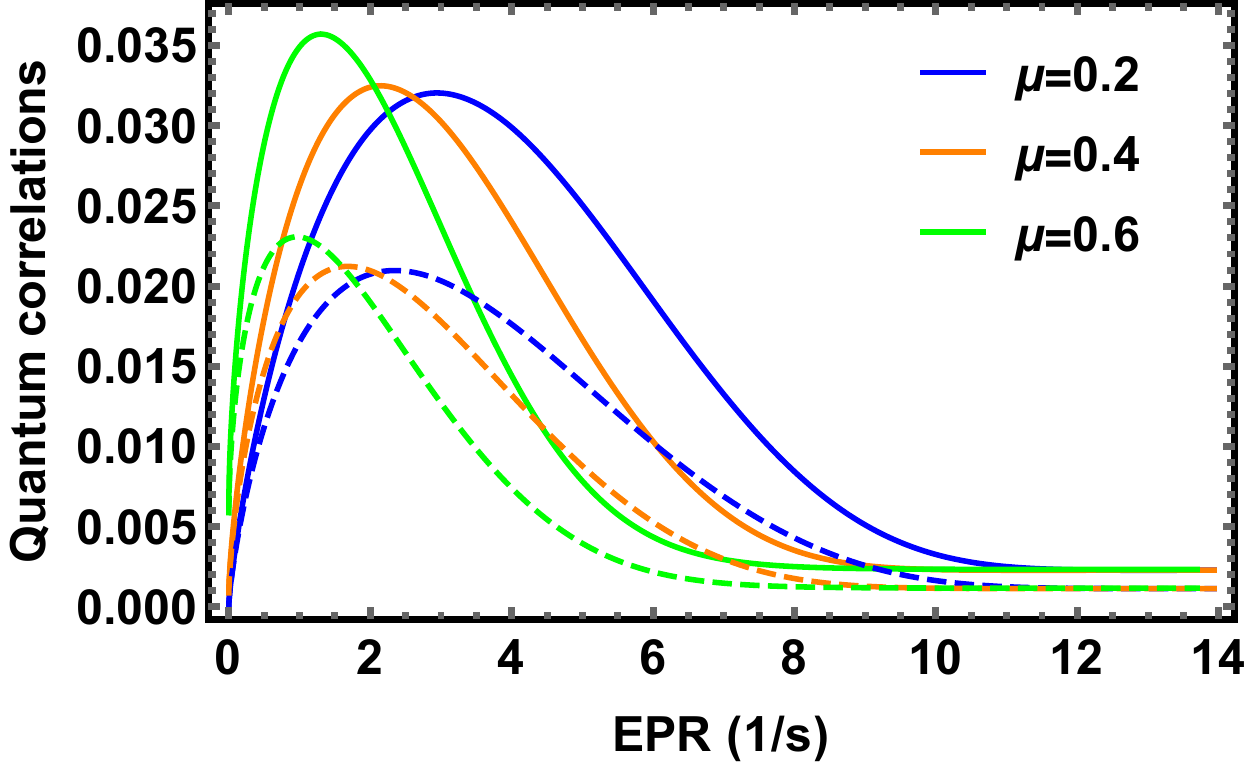}
\includegraphics[width=0.31 \textwidth]{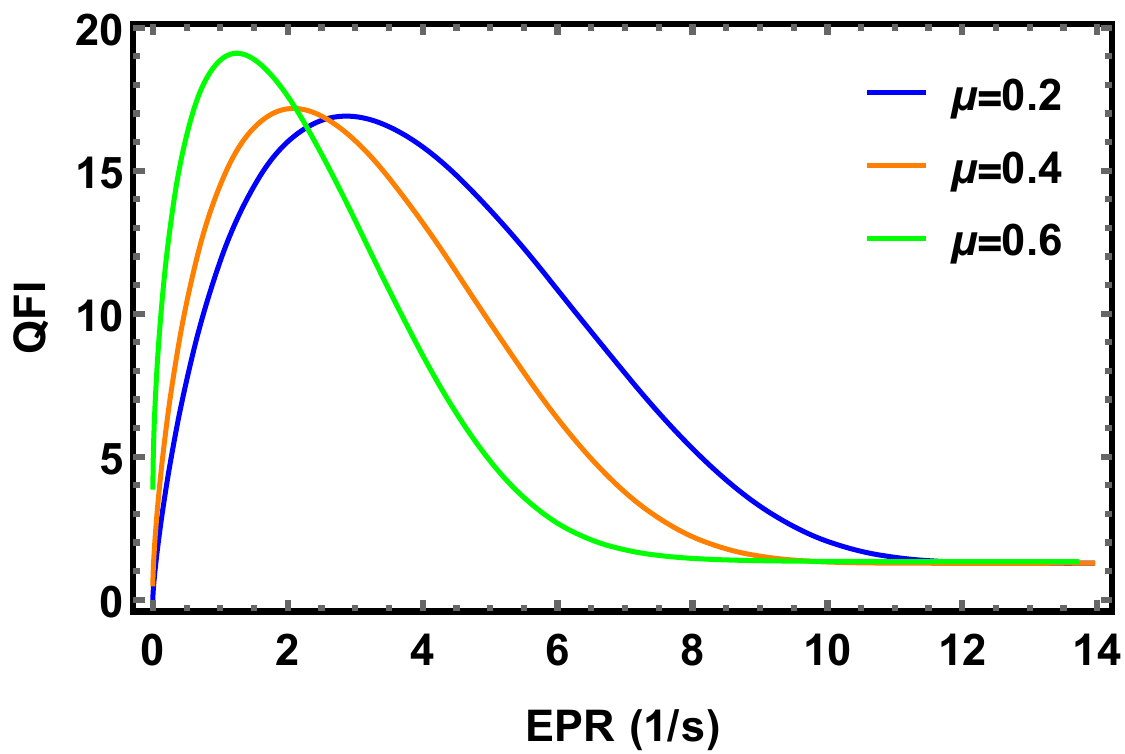}
\caption{Quantum correlations and QFI against EPR for strong tunneling scenario. $\Delta=0.1$. (a) Quantum entanglement against EPR. (b) QMI (solid lines) and QD (dashed lines) vs EPR. (c) QFI vs EPR. For (b) and (c), $\Delta=0.05$. For all, $\mu_1=\mu+\Delta\mu$, $\mu_2=\mu$, $T_1=T_2=0.1$, and $\Gamma_1=\Gamma_2=0.002$, $\omega_1=\omega_2=1$.}
\label{fig.8}
\end{figure}

For the strong tunneling regime, we would have naively expected that the quantum correlation and QFI would be stronger than that in the weak tunneling regime due to a larger tunneling rate. However, this assumption turned out to be over-simplistic. A hint of this over-simplicity can be found from Eq.~\ref{solution}. The coherence in the energy representation $\rho_{23} \approx -i(n_{1m} + n_{2m}) \Delta/2\Gamma$ decays as the tunneling rate becomes higher, which infers that the system is closer to a classical one. Physically, it means that each of the two subsystems is in the steady state with the average of the two reservoirs. Then, the nonequilibrium between the two parties in the system diminishes and the two subsystems are approximately in the same state. In other words, due to the strong coupling, the two systems become one whole system. As can be noted in Fig.~\ref{fig.7} (a), the degree of the nonequilibrium characterized by the temperature bias $\Delta T$ represses the QFI, which is in a sharp contrast with Fig.~\ref{fig.2}. Besides the opposing trends, the magnitude of the QFI is repressed in the strong tunneling regime compared with the weak tunneling case. For the two strongly coupled subsystems (e.g. $\Delta\gg\Gamma$), the nonequilibriumness does not play a significant role in determining the behavior of the system. As mentioned above, the solution merely shows that each subsystem is effectively in contact with both reservoirs simultaneously as a result of the strong coupling between the subsystems and much weaker interactions with the environment. In this case, increasing the nonequilibriumness through the temperature difference is then equivalent to increasing the temperature of the higher temperature reservoir while fixing temperature of the lower temperature bath, or increasing the average temperature of the two baths. This situation is essentially very similar to the equilibrium case of the system in an average temperature reservoir. Therefore, the behaviors of QFI and quantum correlations are expected to be similar to the equilibrium case when the average temperature increases. This is justified from the V-shape light band in Fig.~\ref{fig.7} (b). The QFI reaches the optimized values when one of the chemical potentials is near the resonance with the system, and it diminishes when both chemical potentials are away from the resonance similar to the behavior of the whole system immersed simultaneously in the two baths.

In the strong tunneling regime, a higher thermodynamic cost does not guarantee a higher precision in measurement nor stronger quantum correlations. In the regime of the weak tunneling, the entanglement of the two subsystems vanishes almost identically at finite temperatures due to the mixed state nature of the system. For the large tunneling case, a weak entanglement appears in the system at low temperatures and it shares the similar trend against the increase of the EPR with that of the QD and QMI [Fig.\ref{fig.8} (a,b)]. The similar behavior is observed in the coherence and linear entropy $S_L$. This mimics the average effect of the two reservoirs. When the tunnelling rate is far above the decay rate of the system, the two subsystems essentially becomes one whole system immersed in the two reservoirs simultaneously. When tuning up the bias, the average of the two reservoirs changes correspondingly. This not only has impact on the EPR and currents but also on the quantum correlations and the QFI. This average effect results in the nonzero values of the quantum correlations and QFI at zero EPR (e.g. for $\mu=0.6$ case in Fig.~\ref{fig.8}), which is in sharp contrast with Fig.~\ref{fig.4} and \ref{fig.5}, where no quantum correlations or QFI is witnessed at zero EPR and that significantly stronger correlations and QFI's are observed at large EPR's.

\bigskip
\section{Remarks and Conclusion}
In this work, we study the quantum metrology and quantum correlations of two-fermionic systems in nonequilibrium environments. The relationship between them and the thermodynamic cost (EPR) is investigated. The meaning of the QFI in our study is how much data about the tunneling rate a measurement can in principle obtain from the system allowing any POVM measurement. For the parameter estimation of the tunneling rate $\Delta$, we remark that in the nonequilibrium case, an enhancement of the QFI is seen away from the resonant frequencies when the tunneling rate is low. By amplifying the energy current and the EPR, a universal enhancement of the QFI is witnessed. From a more practical perspective, the weak inter-site tunneling is in the regime where the parameter is hard to measure accurately from an experiment, and it is a more relevant regime for realistic consideration. On the other hand, for a large tunneling rate the nonequilibrium conditions do not boost the quantum metrology and correlations significantly, and this regime is less important for our consideration. To conclude, in this study we propose a new approach which can be potentially applied to improve the quantum information measures and correlations in an open system. In this approach, entropy production or thermodynamic cost is needed and can be exploited to achieve the desired enhancement. We also pointed out that the nonequilibrium enhancement comes at the thermodynamic cost of energy consumption and dissipation from the external sources, which might trade off some nonequilibrium enhancement by introducing extra disorder to the controlled environments depending on the specifics of the experiments. 

The analyses and results in this paper may provide insights for the study of quantum metrology and related fields. Though the experimental verification is yet to be performed, the general method proposed in this study can be helpful in guiding the design of certain future experiments. Furthermore, for the basic understandings of nonequilibrium quantum phenomena, the relationship between the thermodynamic cost and the quantum mechanical nature of a system characterized by the Fisher information and the quantum correlations has been investigated but not fully understood, and how to enhance them in an open system is an important question in many areas of studies. The equilibrium scenario is relatively well-understood, but the physics of far-from-equilibrium scenario remains elusive. We hope that our study can shed light on the understanding of nonequilibrium physics and its application in quantum metrology and quantum information sciences.

The entropy production in this work is treated in an approximate and semi-classical way. More careful analysis can be done to incorporate the corrections of quantum thermodynamics. In addition, the environment is treated as classical and the Markovian approximation is assumed. The method and conclusions drawn from this model can be directly extended to the study the quantum processor based on dimer qubit \cite{petrosyan2002scalable, petrosyan2006quantum}, wherein two similar two-level systems separated by a few nanometers and interacting via resonant dipole-dipole interactions as well as general bipartite two-level open quantum systems  \cite{wang2018coherence}. 

\bigskip
\begin{acknowledgments}
X.W wants to thank Kun Zhang for numerous helpful discussions.
\end{acknowledgments}
\bigskip
\appendix

\section{Solution to the quantum master equation}\label{append.a}
The quantum master equation with the above approximations can be shown to have the following form,
\beq \dot \rho_S(t)=i[\rho_S,H_S]-D_0[\rho]-D_s[\rho], \label{mst} \eeq
where the dissipation terms are
\beq D_0[\rho]=\sum_{i=1}^2N_i[\rho], \ \ D_s[\rho]=\sum_{i=1}^2 S_i[\rho]\, . \eeq
Without constraining the specific properties of the reservoirs, $N_i[\rho]$ and $S_i[\rho]$ are defined as follows (plus and minus signs are for bosonic reservoirs and fermionic reservoirs, respectively),
\begin{widetext}
\begin{align} 
N_i[\rho]&=\Gamma_1 \cdot \frac{1}{2}[1+(-1)^i \cos \theta] \ \left[(1\pm n_1^{T_i})(\zeta_1^\dagger \zeta_1 \tilde \rho - \zeta_1 \rhot \zeta_1^\dagger) +n_1^{T_i} (\zeta_1 \zeta_1^\dagger \rhot-\zeta_1^\dagger \rhot \zeta_1)+h.c. \right] \nonumber\\
&+\Gamma_2 \cdot \frac{1}{2}[1+(-1)^{i-1} \cos \theta] \ \left[(1\pm n_2^{T_i})(\zeta_2^\dagger \zeta_2 \tilde \rho - \zeta_2 \rhot \zeta_2^\dagger) +n_2^{T_i} (\zeta_2 \zeta_2^\dagger \rhot-\zeta_2^\dagger \rhot \zeta_2)+h.c.\right],\end{align}
and
\begin{align}
S_i[\rho]&=(-1)^{i-1} \frac{1}{2} \Gamma_1 \sin \theta \ \left[(1 \pm n_1^{T_i})(\zeta_2^\dagger \zeta_1 \rhot  -  \zeta_1 \rhot \zeta_2^\dagger) + n_1^{T_i} (\zeta_2 \zeta_1^\dagger \rhot- \zeta_1^\dagger \rhot \zeta_2 )+h.c.\right] \nonumber\\
& + (-1)^{i-1} \frac{1}{2} \Gamma_2 \sin\theta \ \left[(1 \pm n_2^{T_i})(\zeta_1^\dagger \zeta_2 \rhot  - \zeta_1 \rhot \zeta_2^\dagger ) +n_2^{T_i} (\zeta_1 \zeta_2^\dagger \rhot - \zeta_1^\dagger \rhot \zeta_2 )+h.c. \right] \,.
\end{align} \end{widetext}
where $\omega_a'$ is the energy eigenvalue of the system, $\sin\theta$ and $\cos\theta$ are the elements in the transformation matrix defined in \ref{newH}, and $n_k^{T_i}$ is the number density of the $i^{\rm{th}}$ reservoir with temperature $T$ and energy $\omega^\prime_k$. The dissipator $D_0$ describes the particle exchanges with the reservoirs, and $D_s$ gives the coherence between energy levels of the system which is absent in the Lindblad formalism and important for studies of quantum systems in nonequilibrium environments.

Due to the rapid oscillation of field modes, we assume the Weisskopf-Wigner approximation. We expand the time integral to infinity and replace the summation in the interaction Hamiltonian by integration. The decay rates after the approximation are defined as follows,
\beq \Gamma_i \equiv \frac{2V}{(2\pi)^3}\pi\int d^3\vec{k}\ \lambda^2_{\vec{k}}\ \delta(\omega'_i-\omega_k). \eeq
In this study, we assume that the number density of the reservoir is the standard Fermi-Dirac distribution, $n_k^{T_i}=\dfrac{1}{e^{\beta_i(\omega'_k-\mu_i)}+1}$.

\section{Positivity of the EPR formula}\label{positivity}
The master equation and our definition of currents are coarse-grained descriptions of the quantum system. The approach is an approximation which under certain limits may appear to break the classical laws of thermodynamics. Therefore, the range of applicability is an issue that needs to be verified. Here, we provide a simple proof that this definition is valid in the parameter regime discussed in this paper. The particle current from bath $l$ is defined as the change of the number of system particles per unit time due to the existence of the bath $l$. The simplified expression from Eq.~\ref{parcurnt} is:
\begin{equation}
    I_l = \tr(D_l[\rho_S]\mathcal{N}_S)\, .
\end{equation}
Similarly, the energy current from the bath $l$ is:
\begin{equation}
    J_l = \tr(D_l[\rho_S]\calh_S)\,.
\end{equation}
Assuming the symmetric sites $\omega_1=\omega_2$ and that the inter-site tunneling rate is much lower than the system frequency $\Delta\ll \omega$, we apply the leading order solution \eqref{solution} to the semi-classical expression of the EPR \eqref{epr}. The entropy production is defined classically by $TdS=dU-\mu dN$. The EPR can be approximately expressed as follows:
\begin{align}
    \dot S&=-J_1\left(\frac{1}{T_1}-\frac{1}{T_2}\right)+I_1\left(\frac{\mu_1}{T_1}-\frac{\mu_2}{T_2}\right)\\
    &\simeq \omega(\beta_1-\beta_2)\left((1-n_1)(\rho_{11}+\rho_{44})+n_1(\rho_{11}+\rho_{33})+(1-n_1)(\rho_{33}+\rho_{44})+n_1(\rho_{11}+\rho_{22})\right) \\
   &\quad -(\mu_1\beta_1-\mu_2\beta_2)\left((1-n_1)(\rho_{22}+\rho_{44})-n_1(\rho_{11}+\rho_{33})+(1-n_1)(\rho_{33}+\rho_{44})-n_1(\rho_{11}+\rho_{22})\right)\,.
\end{align}
The leading order elements of the density matrix are given in Eq.\eqref{solution}. The steady state solution of the master equation results in the following EPR:
\begin{align}
    \dot S \simeq (\mu_1\beta_1+\omega\beta_2-\mu_2\beta_2-\omega\beta_1)&(2e^{2\mu_1\beta_1+(\mu_2+\omega)\beta_2}+2e^{\mu_1\beta_1+\omega\beta_1+2\omega\beta_2}-2e^{2\mu_2\beta_2+(\mu_1+\omega)\beta_1}\\
    &-2e^{\mu_2\beta_2+\beta_1\omega+(\beta_1+\beta_2)\omega}+2e^{2\mu_1\beta_1+2\omega\beta_2}-2e^{2\mu_2\beta_2+2 \omega\beta_1})f(\mu_i,\beta_i)\,,
\end{align}
where $f(\mu_i,\beta_i)$ is a positive function defined as follows:
\begin{align}
    f^{-1}(\mu_i,\beta_i)=2(e^{\mu_1\beta_1}+e^{\omega\beta_1})^2(e^{\mu_2\beta_2}+e^{\omega\beta_2})^2/(\beta_1\beta_2)\,.
\end{align}
Since we are only interested in the issue of positivity, we will ignore this positive factor. Regrouping the expression of EPR, we can write it as follows:
\begin{eqnarray}
    \dot S\simeq & f(\mu_i,\beta_i)\cdot \big[(\mu_1\beta_1+\omega\beta_2)-(\mu_2\beta_2+\omega\beta_1)\big]\cdot \Big[ (2e^{2\mu_1\beta_1+(\mu_2+\omega)\beta_2}-2e^{2\mu_2\beta_2+(\mu_1+\omega)\beta_1})+(2e^{\mu_1\beta_1+\omega\beta_1+2\omega\beta_2}\nonumber\\
    &-2e^{\mu_2\beta_2+\beta_1\omega+(\beta_1+\beta_2)\omega})+(2e^{2\mu_1\beta_1+2\omega\beta_2}-2e^{2\mu_2\beta_2+2 \omega\beta_1})\Big]\nonumber\\
    =& f(\mu_i,\beta_i)\cdot \big[(\mu_1\beta_1+\omega\beta_2)-(\mu_2\beta_2+\omega\beta_1)\big]\cdot \Big[ e^{\mu_1\beta_1+\mu_2\beta_2}(2e^{\mu_1\beta_1+\omega\beta_2}-2e^{\mu_2\beta_2+\omega\beta_1})+e^{\omega\beta_1+\omega\beta_2}(2e^{\mu_1\beta_1+\omega\beta_2}\nonumber\\
    &-2e^{\mu_2\beta_2+\omega\beta_1})+(2e^{2\mu_1\beta_1+2\omega\beta_2}-2e^{2\mu_2\beta_2+2 \omega\beta_1})\Big]
\end{eqnarray}
We notice that in the last equation, each term in the parentheses of the second square bracket has the same sign as the first square bracket $\big[(\mu_1\beta_1+\omega\beta_2)-(\mu_2\beta_2+\omega\beta_1)\big]$. Therefore, the whole expression is non-negative. This derivation suggests that in the regimes we are interested in, the semi-classical expression for the entropy production is well-behaved and we can assume that it gives the reasonably accurate value for the classical EPR of our system.

\bigskip
\providecommand{\noopsort}[1]{}\providecommand{\singleletter}[1]{#1}%


\begin{thebibliography}{76}%
\makeatletter
\providecommand \@ifxundefined [1]{%
 \@ifx{#1\undefined}
}%
\providecommand \@ifnum [1]{%
 \ifnum #1\expandafter \@firstoftwo
 \else \expandafter \@secondoftwo
 \fi
}%
\providecommand \@ifx [1]{%
 \ifx #1\expandafter \@firstoftwo
 \else \expandafter \@secondoftwo
 \fi
}%
\providecommand \natexlab [1]{#1}%
\providecommand \enquote  [1]{``#1''}%
\providecommand \bibnamefont  [1]{#1}%
\providecommand \bibfnamefont [1]{#1}%
\providecommand \citenamefont [1]{#1}%
\providecommand \href@noop [0]{\@secondoftwo}%
\providecommand \href [0]{\begingroup \@sanitize@url \@href}%
\providecommand \@href[1]{\@@startlink{#1}\@@href}%
\providecommand \@@href[1]{\endgroup#1\@@endlink}%
\providecommand \@sanitize@url [0]{\catcode `\\12\catcode `\$12\catcode
  `\&12\catcode `\#12\catcode `\^12\catcode `\_12\catcode `\%12\relax}%
\providecommand \@@startlink[1]{}%
\providecommand \@@endlink[0]{}%
\providecommand \url  [0]{\begingroup\@sanitize@url \@url }%
\providecommand \@url [1]{\endgroup\@href {#1}{\urlprefix }}%
\providecommand \urlprefix  [0]{URL }%
\providecommand \Eprint [0]{\href }%
\providecommand \doibase [0]{https://doi.org/}%
\providecommand \selectlanguage [0]{\@gobble}%
\providecommand \bibinfo  [0]{\@secondoftwo}%
\providecommand \bibfield  [0]{\@secondoftwo}%
\providecommand \translation [1]{[#1]}%
\providecommand \BibitemOpen [0]{}%
\providecommand \bibitemStop [0]{}%
\providecommand \bibitemNoStop [0]{.\EOS\space}%
\providecommand \EOS [0]{\spacefactor3000\relax}%
\providecommand \BibitemShut  [1]{\csname bibitem#1\endcsname}%
\let\auto@bib@innerbib\@empty
\bibitem [{\citenamefont {Nolan}\ \emph {et~al.}(2016)\citenamefont {Nolan},
  \citenamefont {Sabbatini}, \citenamefont {Bromley}, \citenamefont {Davis},\
  and\ \citenamefont {Haine}}]{nolan2016quantum}%
  \BibitemOpen
  \bibfield  {author} {\bibinfo {author} {\bibfnamefont {S.~P.}\ \bibnamefont
  {Nolan}}, \bibinfo {author} {\bibfnamefont {J.}~\bibnamefont {Sabbatini}},
  \bibinfo {author} {\bibfnamefont {M.~W.}\ \bibnamefont {Bromley}}, \bibinfo
  {author} {\bibfnamefont {M.~J.}\ \bibnamefont {Davis}},\ and\ \bibinfo
  {author} {\bibfnamefont {S.~A.}\ \bibnamefont {Haine}},\ }\bibfield  {title}
  {\bibinfo {title} {Quantum enhanced measurement of rotations with a spin-1
  bose-einstein condensate in a ring trap},\ }\href@noop {} {\bibfield
  {journal} {\bibinfo  {journal} {Physical Review A}\ }\textbf {\bibinfo
  {volume} {93}},\ \bibinfo {pages} {023616} (\bibinfo {year}
  {2016})}\BibitemShut {NoStop}%
\bibitem [{\citenamefont {Cooper}\ \emph {et~al.}(2012)\citenamefont {Cooper},
  \citenamefont {Hallwood}, \citenamefont {Dunningham},\ and\ \citenamefont
  {Brand}}]{cooper2012robust}%
  \BibitemOpen
  \bibfield  {author} {\bibinfo {author} {\bibfnamefont {J.}~\bibnamefont
  {Cooper}}, \bibinfo {author} {\bibfnamefont {D.}~\bibnamefont {Hallwood}},
  \bibinfo {author} {\bibfnamefont {J.}~\bibnamefont {Dunningham}},\ and\
  \bibinfo {author} {\bibfnamefont {J.}~\bibnamefont {Brand}},\ }\bibfield
  {title} {\bibinfo {title} {Robust quantum enhanced phase estimation in a
  multimode interferometer},\ }\href@noop {} {\bibfield  {journal} {\bibinfo
  {journal} {Physical review letters}\ }\textbf {\bibinfo {volume} {108}},\
  \bibinfo {pages} {130402} (\bibinfo {year} {2012})}\BibitemShut {NoStop}%
\bibitem [{\citenamefont {Giovannetti}\ \emph {et~al.}(2004)\citenamefont
  {Giovannetti}, \citenamefont {Lloyd},\ and\ \citenamefont
  {Maccone}}]{giovannetti2004quantum}%
  \BibitemOpen
  \bibfield  {author} {\bibinfo {author} {\bibfnamefont {V.}~\bibnamefont
  {Giovannetti}}, \bibinfo {author} {\bibfnamefont {S.}~\bibnamefont {Lloyd}},\
  and\ \bibinfo {author} {\bibfnamefont {L.}~\bibnamefont {Maccone}},\
  }\bibfield  {title} {\bibinfo {title} {Quantum-enhanced measurements: beating
  the standard quantum limit},\ }\href@noop {} {\bibfield  {journal} {\bibinfo
  {journal} {Science}\ }\textbf {\bibinfo {volume} {306}},\ \bibinfo {pages}
  {1330} (\bibinfo {year} {2004})}\BibitemShut {NoStop}%
\bibitem [{\citenamefont {Wolfgramm}\ \emph {et~al.}(2013)\citenamefont
  {Wolfgramm}, \citenamefont {Vitelli}, \citenamefont {Beduini}, \citenamefont
  {Godbout},\ and\ \citenamefont {Mitchell}}]{wolfgramm2013entanglement}%
  \BibitemOpen
  \bibfield  {author} {\bibinfo {author} {\bibfnamefont {F.}~\bibnamefont
  {Wolfgramm}}, \bibinfo {author} {\bibfnamefont {C.}~\bibnamefont {Vitelli}},
  \bibinfo {author} {\bibfnamefont {F.~A.}\ \bibnamefont {Beduini}}, \bibinfo
  {author} {\bibfnamefont {N.}~\bibnamefont {Godbout}},\ and\ \bibinfo {author}
  {\bibfnamefont {M.~W.}\ \bibnamefont {Mitchell}},\ }\bibfield  {title}
  {\bibinfo {title} {Entanglement-enhanced probing of a delicate material
  system},\ }\href@noop {} {\bibfield  {journal} {\bibinfo  {journal} {Nature
  Photonics}\ }\textbf {\bibinfo {volume} {7}},\ \bibinfo {pages} {28}
  (\bibinfo {year} {2013})}\BibitemShut {NoStop}%
\bibitem [{\citenamefont {Yonezawa}\ \emph {et~al.}(2012)\citenamefont
  {Yonezawa}, \citenamefont {Nakane}, \citenamefont {Wheatley}, \citenamefont
  {Iwasawa}, \citenamefont {Takeda}, \citenamefont {Arao}, \citenamefont
  {Ohki}, \citenamefont {Tsumura}, \citenamefont {Berry}, \citenamefont {Ralph}
  \emph {et~al.}}]{yonezawa2012quantum}%
  \BibitemOpen
  \bibfield  {author} {\bibinfo {author} {\bibfnamefont {H.}~\bibnamefont
  {Yonezawa}}, \bibinfo {author} {\bibfnamefont {D.}~\bibnamefont {Nakane}},
  \bibinfo {author} {\bibfnamefont {T.~A.}\ \bibnamefont {Wheatley}}, \bibinfo
  {author} {\bibfnamefont {K.}~\bibnamefont {Iwasawa}}, \bibinfo {author}
  {\bibfnamefont {S.}~\bibnamefont {Takeda}}, \bibinfo {author} {\bibfnamefont
  {H.}~\bibnamefont {Arao}}, \bibinfo {author} {\bibfnamefont {K.}~\bibnamefont
  {Ohki}}, \bibinfo {author} {\bibfnamefont {K.}~\bibnamefont {Tsumura}},
  \bibinfo {author} {\bibfnamefont {D.~W.}\ \bibnamefont {Berry}}, \bibinfo
  {author} {\bibfnamefont {T.~C.}\ \bibnamefont {Ralph}}, \emph {et~al.},\
  }\bibfield  {title} {\bibinfo {title} {Quantum-enhanced optical-phase
  tracking},\ }\href@noop {} {\bibfield  {journal} {\bibinfo  {journal}
  {Science}\ }\textbf {\bibinfo {volume} {337}},\ \bibinfo {pages} {1514}
  (\bibinfo {year} {2012})}\BibitemShut {NoStop}%
\bibitem [{\citenamefont {Braun}\ \emph {et~al.}(2018)\citenamefont {Braun},
  \citenamefont {Adesso}, \citenamefont {Benatti}, \citenamefont {Floreanini},
  \citenamefont {Marzolino}, \citenamefont {Mitchell},\ and\ \citenamefont
  {Pirandola}}]{braun2018quantum}%
  \BibitemOpen
  \bibfield  {author} {\bibinfo {author} {\bibfnamefont {D.}~\bibnamefont
  {Braun}}, \bibinfo {author} {\bibfnamefont {G.}~\bibnamefont {Adesso}},
  \bibinfo {author} {\bibfnamefont {F.}~\bibnamefont {Benatti}}, \bibinfo
  {author} {\bibfnamefont {R.}~\bibnamefont {Floreanini}}, \bibinfo {author}
  {\bibfnamefont {U.}~\bibnamefont {Marzolino}}, \bibinfo {author}
  {\bibfnamefont {M.~W.}\ \bibnamefont {Mitchell}},\ and\ \bibinfo {author}
  {\bibfnamefont {S.}~\bibnamefont {Pirandola}},\ }\bibfield  {title} {\bibinfo
  {title} {Quantum-enhanced measurements without entanglement},\ }\href@noop {}
  {\bibfield  {journal} {\bibinfo  {journal} {Reviews of Modern Physics}\
  }\textbf {\bibinfo {volume} {90}},\ \bibinfo {pages} {035006} (\bibinfo
  {year} {2018})}\BibitemShut {NoStop}%
\bibitem [{\citenamefont {Yurke}(1986)}]{yurke1986input}%
  \BibitemOpen
  \bibfield  {author} {\bibinfo {author} {\bibfnamefont {B.}~\bibnamefont
  {Yurke}},\ }\bibfield  {title} {\bibinfo {title} {Input states for
  enhancement of fermion interferometer sensitivity},\ }\href@noop {}
  {\bibfield  {journal} {\bibinfo  {journal} {Physical review letters}\
  }\textbf {\bibinfo {volume} {56}},\ \bibinfo {pages} {1515} (\bibinfo {year}
  {1986})}\BibitemShut {NoStop}%
\bibitem [{\citenamefont {Scheff{\'e}}\ \emph {et~al.}(1947)\citenamefont
  {Scheff{\'e}} \emph {et~al.}}]{scheffe1947h}%
  \BibitemOpen
  \bibfield  {author} {\bibinfo {author} {\bibfnamefont {H.}~\bibnamefont
  {Scheff{\'e}}} \emph {et~al.},\ }\bibfield  {title} {\bibinfo {title} {H.
  cram{\'e}r, mathematical methods of statistics},\ }\href@noop {} {\bibfield
  {journal} {\bibinfo  {journal} {Bulletin of the American Mathematical
  Society}\ }\textbf {\bibinfo {volume} {53}},\ \bibinfo {pages} {733}
  (\bibinfo {year} {1947})}\BibitemShut {NoStop}%
\bibitem [{\citenamefont {Braunstein}\ and\ \citenamefont
  {Caves}(1994)}]{braunstein1994statistical}%
  \BibitemOpen
  \bibfield  {author} {\bibinfo {author} {\bibfnamefont {S.~L.}\ \bibnamefont
  {Braunstein}}\ and\ \bibinfo {author} {\bibfnamefont {C.~M.}\ \bibnamefont
  {Caves}},\ }\bibfield  {title} {\bibinfo {title} {Statistical distance and
  the geometry of quantum states},\ }\href@noop {} {\bibfield  {journal}
  {\bibinfo  {journal} {Physical Review Letters}\ }\textbf {\bibinfo {volume}
  {72}},\ \bibinfo {pages} {3439} (\bibinfo {year} {1994})}\BibitemShut
  {NoStop}%
\bibitem [{\citenamefont {Uhlmann}(1991)}]{uhlmann1991gauge}%
  \BibitemOpen
  \bibfield  {author} {\bibinfo {author} {\bibfnamefont {A.}~\bibnamefont
  {Uhlmann}},\ }\bibfield  {title} {\bibinfo {title} {A gauge field governing
  parallel transport along mixed states},\ }\href@noop {} {\bibfield  {journal}
  {\bibinfo  {journal} {letters in mathematical physics}\ }\textbf {\bibinfo
  {volume} {21}},\ \bibinfo {pages} {229} (\bibinfo {year} {1991})}\BibitemShut
  {NoStop}%
\bibitem [{\citenamefont {Luo}\ and\ \citenamefont
  {Zhang}(2004)}]{luo2004informational}%
  \BibitemOpen
  \bibfield  {author} {\bibinfo {author} {\bibfnamefont {S.}~\bibnamefont
  {Luo}}\ and\ \bibinfo {author} {\bibfnamefont {Q.}~\bibnamefont {Zhang}},\
  }\bibfield  {title} {\bibinfo {title} {Informational distance on
  quantum-state space},\ }\href@noop {} {\bibfield  {journal} {\bibinfo
  {journal} {Physical Review A}\ }\textbf {\bibinfo {volume} {69}},\ \bibinfo
  {pages} {032106} (\bibinfo {year} {2004})}\BibitemShut {NoStop}%
\bibitem [{\citenamefont {Marzolino}\ and\ \citenamefont
  {Prosen}(2017)}]{marzolino2017fisher}%
  \BibitemOpen
  \bibfield  {author} {\bibinfo {author} {\bibfnamefont {U.}~\bibnamefont
  {Marzolino}}\ and\ \bibinfo {author} {\bibfnamefont {T.}~\bibnamefont
  {Prosen}},\ }\bibfield  {title} {\bibinfo {title} {Fisher information
  approach to nonequilibrium phase transitions in a quantum xxz spin chain with
  boundary noise},\ }\href@noop {} {\bibfield  {journal} {\bibinfo  {journal}
  {Physical Review B}\ }\textbf {\bibinfo {volume} {96}},\ \bibinfo {pages}
  {104402} (\bibinfo {year} {2017})}\BibitemShut {NoStop}%
\bibitem [{\citenamefont {Banerjee}\ \emph {et~al.}(2018)\citenamefont
  {Banerjee}, \citenamefont {Erdmenger},\ and\ \citenamefont
  {Sarkar}}]{banerjee2018connecting}%
  \BibitemOpen
  \bibfield  {author} {\bibinfo {author} {\bibfnamefont {S.}~\bibnamefont
  {Banerjee}}, \bibinfo {author} {\bibfnamefont {J.}~\bibnamefont
  {Erdmenger}},\ and\ \bibinfo {author} {\bibfnamefont {D.}~\bibnamefont
  {Sarkar}},\ }\bibfield  {title} {\bibinfo {title} {Connecting fisher
  information to bulk entanglement in holography},\ }\href@noop {} {\bibfield
  {journal} {\bibinfo  {journal} {Journal of High Energy Physics}\ }\textbf
  {\bibinfo {volume} {2018}},\ \bibinfo {pages} {1} (\bibinfo {year}
  {2018})}\BibitemShut {NoStop}%
\bibitem [{\citenamefont {Sarkar}\ \emph {et~al.}(2017)\citenamefont {Sarkar},
  \citenamefont {Banerjee},\ and\ \citenamefont
  {Erdmenger}}]{sarkar2017holographic}%
  \BibitemOpen
  \bibfield  {author} {\bibinfo {author} {\bibfnamefont {D.}~\bibnamefont
  {Sarkar}}, \bibinfo {author} {\bibfnamefont {S.}~\bibnamefont {Banerjee}},\
  and\ \bibinfo {author} {\bibfnamefont {J.}~\bibnamefont {Erdmenger}},\
  }\bibfield  {title} {\bibinfo {title} {A holographic dual to fisher
  information and its relation with bulk entanglement},\ }in\ \href@noop {}
  {\emph {\bibinfo {booktitle} {Corfu Summer Institute 2016" School and
  Workshops on Elementary Particle Physics and Gravity"}}},\ Vol.\ \bibinfo
  {volume} {292}\ (\bibinfo {organization} {SISSA Medialab},\ \bibinfo {year}
  {2017})\ p.\ \bibinfo {pages} {092}\BibitemShut {NoStop}%
\bibitem [{\citenamefont {Lashkari}\ and\ \citenamefont
  {Van~Raamsdonk}(2016)}]{lashkari2016canonical}%
  \BibitemOpen
  \bibfield  {author} {\bibinfo {author} {\bibfnamefont {N.}~\bibnamefont
  {Lashkari}}\ and\ \bibinfo {author} {\bibfnamefont {M.}~\bibnamefont
  {Van~Raamsdonk}},\ }\bibfield  {title} {\bibinfo {title} {Canonical energy is
  quantum fisher information},\ }\href@noop {} {\bibfield  {journal} {\bibinfo
  {journal} {Journal of High Energy Physics}\ }\textbf {\bibinfo {volume}
  {2016}},\ \bibinfo {pages} {153} (\bibinfo {year} {2016})}\BibitemShut
  {NoStop}%
\bibitem [{\citenamefont {Ollivier}\ and\ \citenamefont
  {Zurek}(2001)}]{ollivier2001quantum}%
  \BibitemOpen
  \bibfield  {author} {\bibinfo {author} {\bibfnamefont {H.}~\bibnamefont
  {Ollivier}}\ and\ \bibinfo {author} {\bibfnamefont {W.~H.}\ \bibnamefont
  {Zurek}},\ }\bibfield  {title} {\bibinfo {title} {Quantum discord: a measure
  of the quantumness of correlations},\ }\href@noop {} {\bibfield  {journal}
  {\bibinfo  {journal} {Physical review letters}\ }\textbf {\bibinfo {volume}
  {88}},\ \bibinfo {pages} {017901} (\bibinfo {year} {2001})}\BibitemShut
  {NoStop}%
\bibitem [{\citenamefont {Einstein}\ \emph {et~al.}(1935)\citenamefont
  {Einstein}, \citenamefont {Podolsky},\ and\ \citenamefont
  {Rosen}}]{einstein1935can}%
  \BibitemOpen
  \bibfield  {author} {\bibinfo {author} {\bibfnamefont {A.}~\bibnamefont
  {Einstein}}, \bibinfo {author} {\bibfnamefont {B.}~\bibnamefont {Podolsky}},\
  and\ \bibinfo {author} {\bibfnamefont {N.}~\bibnamefont {Rosen}},\ }\bibfield
   {title} {\bibinfo {title} {Can quantum-mechanical description of physical
  reality be considered complete?},\ }\href@noop {} {\bibfield  {journal}
  {\bibinfo  {journal} {Physical review}\ }\textbf {\bibinfo {volume} {47}},\
  \bibinfo {pages} {777} (\bibinfo {year} {1935})}\BibitemShut {NoStop}%
\bibitem [{\citenamefont {Nakano}\ \emph {et~al.}(2013)\citenamefont {Nakano},
  \citenamefont {Piani},\ and\ \citenamefont {Adesso}}]{nakano2013negativity}%
  \BibitemOpen
  \bibfield  {author} {\bibinfo {author} {\bibfnamefont {T.}~\bibnamefont
  {Nakano}}, \bibinfo {author} {\bibfnamefont {M.}~\bibnamefont {Piani}},\ and\
  \bibinfo {author} {\bibfnamefont {G.}~\bibnamefont {Adesso}},\ }\bibfield
  {title} {\bibinfo {title} {Negativity of quantumness and its
  interpretations},\ }\href@noop {} {\bibfield  {journal} {\bibinfo  {journal}
  {Physical Review A}\ }\textbf {\bibinfo {volume} {88}},\ \bibinfo {pages}
  {012117} (\bibinfo {year} {2013})}\BibitemShut {NoStop}%
\bibitem [{\citenamefont {Chitambar}\ and\ \citenamefont
  {Gour}(2019)}]{chitambar2019quantum}%
  \BibitemOpen
  \bibfield  {author} {\bibinfo {author} {\bibfnamefont {E.}~\bibnamefont
  {Chitambar}}\ and\ \bibinfo {author} {\bibfnamefont {G.}~\bibnamefont
  {Gour}},\ }\bibfield  {title} {\bibinfo {title} {Quantum resource theories},\
  }\href@noop {} {\bibfield  {journal} {\bibinfo  {journal} {Reviews of Modern
  Physics}\ }\textbf {\bibinfo {volume} {91}},\ \bibinfo {pages} {025001}
  (\bibinfo {year} {2019})}\BibitemShut {NoStop}%
\bibitem [{\citenamefont {Haase}\ \emph {et~al.}(2016)\citenamefont {Haase},
  \citenamefont {Smirne}, \citenamefont {Huelga}, \citenamefont
  {Ko{\l}odynski},\ and\ \citenamefont
  {Demkowicz-Dobrzanski}}]{haase2016precision}%
  \BibitemOpen
  \bibfield  {author} {\bibinfo {author} {\bibfnamefont {J.~F.}\ \bibnamefont
  {Haase}}, \bibinfo {author} {\bibfnamefont {A.}~\bibnamefont {Smirne}},
  \bibinfo {author} {\bibfnamefont {S.}~\bibnamefont {Huelga}}, \bibinfo
  {author} {\bibfnamefont {J.}~\bibnamefont {Ko{\l}odynski}},\ and\ \bibinfo
  {author} {\bibfnamefont {R.}~\bibnamefont {Demkowicz-Dobrzanski}},\
  }\bibfield  {title} {\bibinfo {title} {Precision limits in quantum metrology
  with open quantum systems},\ }\href@noop {} {\bibfield  {journal} {\bibinfo
  {journal} {Quantum Measurements and Quantum Metrology}\ }\textbf {\bibinfo
  {volume} {5}},\ \bibinfo {pages} {13} (\bibinfo {year} {2016})}\BibitemShut
  {NoStop}%
\bibitem [{\citenamefont {T{\'o}th}\ and\ \citenamefont
  {Apellaniz}(2014)}]{toth2014quantum}%
  \BibitemOpen
  \bibfield  {author} {\bibinfo {author} {\bibfnamefont {G.}~\bibnamefont
  {T{\'o}th}}\ and\ \bibinfo {author} {\bibfnamefont {I.}~\bibnamefont
  {Apellaniz}},\ }\bibfield  {title} {\bibinfo {title} {Quantum metrology from
  a quantum information science perspective},\ }\href@noop {} {\bibfield
  {journal} {\bibinfo  {journal} {Journal of Physics A: Mathematical and
  Theoretical}\ }\textbf {\bibinfo {volume} {47}},\ \bibinfo {pages} {424006}
  (\bibinfo {year} {2014})}\BibitemShut {NoStop}%
\bibitem [{\citenamefont {Hyllus}\ \emph {et~al.}(2012)\citenamefont {Hyllus},
  \citenamefont {Laskowski}, \citenamefont {Krischek}, \citenamefont
  {Schwemmer}, \citenamefont {Wieczorek}, \citenamefont {Weinfurter},
  \citenamefont {Pezz{\'e}},\ and\ \citenamefont {Smerzi}}]{hyllus2012fisher}%
  \BibitemOpen
  \bibfield  {author} {\bibinfo {author} {\bibfnamefont {P.}~\bibnamefont
  {Hyllus}}, \bibinfo {author} {\bibfnamefont {W.}~\bibnamefont {Laskowski}},
  \bibinfo {author} {\bibfnamefont {R.}~\bibnamefont {Krischek}}, \bibinfo
  {author} {\bibfnamefont {C.}~\bibnamefont {Schwemmer}}, \bibinfo {author}
  {\bibfnamefont {W.}~\bibnamefont {Wieczorek}}, \bibinfo {author}
  {\bibfnamefont {H.}~\bibnamefont {Weinfurter}}, \bibinfo {author}
  {\bibfnamefont {L.}~\bibnamefont {Pezz{\'e}}},\ and\ \bibinfo {author}
  {\bibfnamefont {A.}~\bibnamefont {Smerzi}},\ }\bibfield  {title} {\bibinfo
  {title} {Fisher information and multiparticle entanglement},\ }\href@noop {}
  {\bibfield  {journal} {\bibinfo  {journal} {Physical Review A}\ }\textbf
  {\bibinfo {volume} {85}},\ \bibinfo {pages} {022321} (\bibinfo {year}
  {2012})}\BibitemShut {NoStop}%
\bibitem [{\citenamefont {Giovannetti}\ \emph {et~al.}(2011)\citenamefont
  {Giovannetti}, \citenamefont {Lloyd},\ and\ \citenamefont
  {Maccone}}]{giovannetti2011advances}%
  \BibitemOpen
  \bibfield  {author} {\bibinfo {author} {\bibfnamefont {V.}~\bibnamefont
  {Giovannetti}}, \bibinfo {author} {\bibfnamefont {S.}~\bibnamefont {Lloyd}},\
  and\ \bibinfo {author} {\bibfnamefont {L.}~\bibnamefont {Maccone}},\
  }\bibfield  {title} {\bibinfo {title} {Advances in quantum metrology},\
  }\href@noop {} {\bibfield  {journal} {\bibinfo  {journal} {Nature photonics}\
  }\textbf {\bibinfo {volume} {5}},\ \bibinfo {pages} {222} (\bibinfo {year}
  {2011})}\BibitemShut {NoStop}%
\bibitem [{\citenamefont {Wiseman}\ and\ \citenamefont
  {Milburn}(2009)}]{wiseman2009quantum}%
  \BibitemOpen
  \bibfield  {author} {\bibinfo {author} {\bibfnamefont {H.~M.}\ \bibnamefont
  {Wiseman}}\ and\ \bibinfo {author} {\bibfnamefont {G.~J.}\ \bibnamefont
  {Milburn}},\ }\href@noop {} {\emph {\bibinfo {title} {Quantum measurement and
  control}}}\ (\bibinfo  {publisher} {Cambridge university press},\ \bibinfo
  {year} {2009})\BibitemShut {NoStop}%
\bibitem [{\citenamefont {Fr{\"o}wis}\ \emph {et~al.}(2019)\citenamefont
  {Fr{\"o}wis}, \citenamefont {Fadel}, \citenamefont {Treutlein}, \citenamefont
  {Gisin},\ and\ \citenamefont {Brunner}}]{frowis2019does}%
  \BibitemOpen
  \bibfield  {author} {\bibinfo {author} {\bibfnamefont {F.}~\bibnamefont
  {Fr{\"o}wis}}, \bibinfo {author} {\bibfnamefont {M.}~\bibnamefont {Fadel}},
  \bibinfo {author} {\bibfnamefont {P.}~\bibnamefont {Treutlein}}, \bibinfo
  {author} {\bibfnamefont {N.}~\bibnamefont {Gisin}},\ and\ \bibinfo {author}
  {\bibfnamefont {N.}~\bibnamefont {Brunner}},\ }\bibfield  {title} {\bibinfo
  {title} {Does large quantum fisher information imply bell correlations?},\
  }\href@noop {} {\bibfield  {journal} {\bibinfo  {journal} {Physical Review
  A}\ }\textbf {\bibinfo {volume} {99}},\ \bibinfo {pages} {040101} (\bibinfo
  {year} {2019})}\BibitemShut {NoStop}%
\bibitem [{\citenamefont {Kim}\ \emph {et~al.}(2018)\citenamefont {Kim},
  \citenamefont {Li}, \citenamefont {Kumar},\ and\ \citenamefont
  {Wu}}]{kim2018characterizing}%
  \BibitemOpen
  \bibfield  {author} {\bibinfo {author} {\bibfnamefont {S.}~\bibnamefont
  {Kim}}, \bibinfo {author} {\bibfnamefont {L.}~\bibnamefont {Li}}, \bibinfo
  {author} {\bibfnamefont {A.}~\bibnamefont {Kumar}},\ and\ \bibinfo {author}
  {\bibfnamefont {J.}~\bibnamefont {Wu}},\ }\bibfield  {title} {\bibinfo
  {title} {Characterizing nonclassical correlations via local quantum fisher
  information},\ }\href@noop {} {\bibfield  {journal} {\bibinfo  {journal}
  {Physical Review A}\ }\textbf {\bibinfo {volume} {97}},\ \bibinfo {pages}
  {032326} (\bibinfo {year} {2018})}\BibitemShut {NoStop}%
\bibitem [{\citenamefont {Przysikezna}\ \emph {et~al.}(2015)\citenamefont
  {Przysikezna}, \citenamefont {Horodecki}, \citenamefont {Horodecki} \emph
  {et~al.}}]{przysikezna2015quantum}%
  \BibitemOpen
  \bibfield  {author} {\bibinfo {author} {\bibfnamefont {A.}~\bibnamefont
  {Przysikezna}}, \bibinfo {author} {\bibfnamefont {M.}~\bibnamefont
  {Horodecki}}, \bibinfo {author} {\bibfnamefont {P.}~\bibnamefont
  {Horodecki}}, \emph {et~al.},\ }\bibfield  {title} {\bibinfo {title} {Quantum
  metrology: Heisenberg limit with bound entanglement},\ }\href@noop {}
  {\bibfield  {journal} {\bibinfo  {journal} {Physical Review A}\ }\textbf
  {\bibinfo {volume} {92}},\ \bibinfo {pages} {062303} (\bibinfo {year}
  {2015})}\BibitemShut {NoStop}%
\bibitem [{\citenamefont {Spietz}\ \emph {et~al.}(2003)\citenamefont {Spietz},
  \citenamefont {Lehnert}, \citenamefont {Siddiqi},\ and\ \citenamefont
  {Schoelkopf}}]{spietz2003primary}%
  \BibitemOpen
  \bibfield  {author} {\bibinfo {author} {\bibfnamefont {L.}~\bibnamefont
  {Spietz}}, \bibinfo {author} {\bibfnamefont {K.}~\bibnamefont {Lehnert}},
  \bibinfo {author} {\bibfnamefont {I.}~\bibnamefont {Siddiqi}},\ and\ \bibinfo
  {author} {\bibfnamefont {R.}~\bibnamefont {Schoelkopf}},\ }\bibfield  {title}
  {\bibinfo {title} {Primary electronic thermometry using the shot noise of a
  tunnel junction},\ }\href@noop {} {\bibfield  {journal} {\bibinfo  {journal}
  {Science}\ }\textbf {\bibinfo {volume} {300}},\ \bibinfo {pages} {1929}
  (\bibinfo {year} {2003})}\BibitemShut {NoStop}%
\bibitem [{\citenamefont {Savukov}\ \emph {et~al.}(2005)\citenamefont
  {Savukov}, \citenamefont {Seltzer}, \citenamefont {Romalis},\ and\
  \citenamefont {Sauer}}]{savukov2005tunable}%
  \BibitemOpen
  \bibfield  {author} {\bibinfo {author} {\bibfnamefont {I.~M.}\ \bibnamefont
  {Savukov}}, \bibinfo {author} {\bibfnamefont {S.}~\bibnamefont {Seltzer}},
  \bibinfo {author} {\bibfnamefont {M.~V.}\ \bibnamefont {Romalis}},\ and\
  \bibinfo {author} {\bibfnamefont {K.}~\bibnamefont {Sauer}},\ }\bibfield
  {title} {\bibinfo {title} {Tunable atomic magnetometer for detection of
  radio-frequency magnetic fields},\ }\href@noop {} {\bibfield  {journal}
  {\bibinfo  {journal} {Physical review letters}\ }\textbf {\bibinfo {volume}
  {95}},\ \bibinfo {pages} {063004} (\bibinfo {year} {2005})}\BibitemShut
  {NoStop}%
\bibitem [{\citenamefont {Marzolino}\ and\ \citenamefont
  {Prosen}(2014)}]{marzolino2014quantum}%
  \BibitemOpen
  \bibfield  {author} {\bibinfo {author} {\bibfnamefont {U.}~\bibnamefont
  {Marzolino}}\ and\ \bibinfo {author} {\bibfnamefont {T.}~\bibnamefont
  {Prosen}},\ }\bibfield  {title} {\bibinfo {title} {Quantum metrology with
  nonequilibrium steady states of quantum spin chains},\ }\href@noop {}
  {\bibfield  {journal} {\bibinfo  {journal} {Physical Review A}\ }\textbf
  {\bibinfo {volume} {90}},\ \bibinfo {pages} {062130} (\bibinfo {year}
  {2014})}\BibitemShut {NoStop}%
\bibitem [{\citenamefont {Wang}\ \emph {et~al.}(2018)\citenamefont {Wang},
  \citenamefont {Wu}, \citenamefont {Cui},\ and\ \citenamefont
  {Wang}}]{wang2018coherence}%
  \BibitemOpen
  \bibfield  {author} {\bibinfo {author} {\bibfnamefont {Z.}~\bibnamefont
  {Wang}}, \bibinfo {author} {\bibfnamefont {W.}~\bibnamefont {Wu}}, \bibinfo
  {author} {\bibfnamefont {G.}~\bibnamefont {Cui}},\ and\ \bibinfo {author}
  {\bibfnamefont {J.}~\bibnamefont {Wang}},\ }\bibfield  {title} {\bibinfo
  {title} {Coherence enhanced quantum metrology in a nonequilibrium optical
  molecule},\ }\href@noop {} {\bibfield  {journal} {\bibinfo  {journal} {New
  Journal of Physics}\ }\textbf {\bibinfo {volume} {20}},\ \bibinfo {pages}
  {033034} (\bibinfo {year} {2018})}\BibitemShut {NoStop}%
\bibitem [{\citenamefont {Yao}\ \emph {et~al.}(2014)\citenamefont {Yao},
  \citenamefont {Xiao}, \citenamefont {Ge}, \citenamefont {Wang},\ and\
  \citenamefont {Sun}}]{yao2014quantum}%
  \BibitemOpen
  \bibfield  {author} {\bibinfo {author} {\bibfnamefont {Y.}~\bibnamefont
  {Yao}}, \bibinfo {author} {\bibfnamefont {X.}~\bibnamefont {Xiao}}, \bibinfo
  {author} {\bibfnamefont {L.}~\bibnamefont {Ge}}, \bibinfo {author}
  {\bibfnamefont {X.-g.}\ \bibnamefont {Wang}},\ and\ \bibinfo {author}
  {\bibfnamefont {C.-p.}\ \bibnamefont {Sun}},\ }\bibfield  {title} {\bibinfo
  {title} {Quantum fisher information in noninertial frames},\ }\href@noop {}
  {\bibfield  {journal} {\bibinfo  {journal} {Physical Review A}\ }\textbf
  {\bibinfo {volume} {89}},\ \bibinfo {pages} {042336} (\bibinfo {year}
  {2014})}\BibitemShut {NoStop}%
\bibitem [{\citenamefont {Paladino}\ \emph {et~al.}(2014)\citenamefont
  {Paladino}, \citenamefont {Galperin}, \citenamefont {Falci},\ and\
  \citenamefont {Altshuler}}]{noiseinf}%
  \BibitemOpen
  \bibfield  {author} {\bibinfo {author} {\bibfnamefont {E.}~\bibnamefont
  {Paladino}}, \bibinfo {author} {\bibfnamefont {Y.}~\bibnamefont {Galperin}},
  \bibinfo {author} {\bibfnamefont {G.}~\bibnamefont {Falci}},\ and\ \bibinfo
  {author} {\bibfnamefont {B.}~\bibnamefont {Altshuler}},\ }\bibfield  {title}
  {\bibinfo {title} {1/f noise: Implications for solid-state quantum
  information},\ }\href@noop {} {\bibfield  {journal} {\bibinfo  {journal}
  {Reviews of Modern Physics}\ }\textbf {\bibinfo {volume} {86}},\ \bibinfo
  {pages} {361} (\bibinfo {year} {2014})}\BibitemShut {NoStop}%
\bibitem [{\citenamefont {Escher}\ \emph {et~al.}(2011)\citenamefont {Escher},
  \citenamefont {de~Matos~Filho},\ and\ \citenamefont
  {Davidovich}}]{escher2011general}%
  \BibitemOpen
  \bibfield  {author} {\bibinfo {author} {\bibfnamefont {B.}~\bibnamefont
  {Escher}}, \bibinfo {author} {\bibfnamefont {R.}~\bibnamefont
  {de~Matos~Filho}},\ and\ \bibinfo {author} {\bibfnamefont {L.}~\bibnamefont
  {Davidovich}},\ }\bibfield  {title} {\bibinfo {title} {General framework for
  estimating the ultimate precision limit in noisy quantum-enhanced
  metrology},\ }\href@noop {} {\bibfield  {journal} {\bibinfo  {journal}
  {Nature Physics}\ }\textbf {\bibinfo {volume} {7}},\ \bibinfo {pages} {406}
  (\bibinfo {year} {2011})}\BibitemShut {NoStop}%
\bibitem [{\citenamefont {Demkowicz-Dobrza{\'n}ski}\ \emph
  {et~al.}(2012)\citenamefont {Demkowicz-Dobrza{\'n}ski}, \citenamefont
  {Ko{\l}ody{\'n}ski},\ and\ \citenamefont
  {Gu{\c{t}}{\u{a}}}}]{demkowicz2012elusive}%
  \BibitemOpen
  \bibfield  {author} {\bibinfo {author} {\bibfnamefont {R.}~\bibnamefont
  {Demkowicz-Dobrza{\'n}ski}}, \bibinfo {author} {\bibfnamefont
  {J.}~\bibnamefont {Ko{\l}ody{\'n}ski}},\ and\ \bibinfo {author}
  {\bibfnamefont {M.}~\bibnamefont {Gu{\c{t}}{\u{a}}}},\ }\bibfield  {title}
  {\bibinfo {title} {The elusive heisenberg limit in quantum-enhanced
  metrology},\ }\href@noop {} {\bibfield  {journal} {\bibinfo  {journal}
  {Nature communications}\ }\textbf {\bibinfo {volume} {3}},\ \bibinfo {pages}
  {1} (\bibinfo {year} {2012})}\BibitemShut {NoStop}%
\bibitem [{\citenamefont {Thoss}\ and\ \citenamefont
  {Evers}(2018)}]{thoss2018perspective}%
  \BibitemOpen
  \bibfield  {author} {\bibinfo {author} {\bibfnamefont {M.}~\bibnamefont
  {Thoss}}\ and\ \bibinfo {author} {\bibfnamefont {F.}~\bibnamefont {Evers}},\
  }\bibfield  {title} {\bibinfo {title} {Perspective: Theory of quantum
  transport in molecular junctions},\ }\href@noop {} {\bibfield  {journal}
  {\bibinfo  {journal} {The Journal of chemical physics}\ }\textbf {\bibinfo
  {volume} {148}},\ \bibinfo {pages} {030901} (\bibinfo {year}
  {2018})}\BibitemShut {NoStop}%
\bibitem [{\citenamefont {Skourtis}\ \emph {et~al.}(2016)\citenamefont
  {Skourtis}, \citenamefont {Liu}, \citenamefont {Antoniou}, \citenamefont
  {Virshup},\ and\ \citenamefont {Beratan}}]{skourtis2016dexter}%
  \BibitemOpen
  \bibfield  {author} {\bibinfo {author} {\bibfnamefont {S.~S.}\ \bibnamefont
  {Skourtis}}, \bibinfo {author} {\bibfnamefont {C.}~\bibnamefont {Liu}},
  \bibinfo {author} {\bibfnamefont {P.}~\bibnamefont {Antoniou}}, \bibinfo
  {author} {\bibfnamefont {A.~M.}\ \bibnamefont {Virshup}},\ and\ \bibinfo
  {author} {\bibfnamefont {D.~N.}\ \bibnamefont {Beratan}},\ }\bibfield
  {title} {\bibinfo {title} {Dexter energy transfer pathways},\ }\href@noop {}
  {\bibfield  {journal} {\bibinfo  {journal} {Proceedings of the National
  Academy of Sciences}\ }\textbf {\bibinfo {volume} {113}},\ \bibinfo {pages}
  {8115} (\bibinfo {year} {2016})}\BibitemShut {NoStop}%
\bibitem [{\citenamefont {Santos}\ \emph {et~al.}(2019)\citenamefont {Santos},
  \citenamefont {C{\'e}leri}, \citenamefont {Landi},\ and\ \citenamefont
  {Paternostro}}]{santos2019role}%
  \BibitemOpen
  \bibfield  {author} {\bibinfo {author} {\bibfnamefont {J.~P.}\ \bibnamefont
  {Santos}}, \bibinfo {author} {\bibfnamefont {L.~C.}\ \bibnamefont
  {C{\'e}leri}}, \bibinfo {author} {\bibfnamefont {G.~T.}\ \bibnamefont
  {Landi}},\ and\ \bibinfo {author} {\bibfnamefont {M.}~\bibnamefont
  {Paternostro}},\ }\bibfield  {title} {\bibinfo {title} {The role of quantum
  coherence in non-equilibrium entropy production},\ }\href@noop {} {\bibfield
  {journal} {\bibinfo  {journal} {npj Quantum Information}\ }\textbf {\bibinfo
  {volume} {5}},\ \bibinfo {pages} {1} (\bibinfo {year} {2019})}\BibitemShut
  {NoStop}%
\bibitem [{\citenamefont {Landauer}(1961)}]{landauer1961irreversibility}%
  \BibitemOpen
  \bibfield  {author} {\bibinfo {author} {\bibfnamefont {R.}~\bibnamefont
  {Landauer}},\ }\bibfield  {title} {\bibinfo {title} {Irreversibility and heat
  generation in the computing process},\ }\href@noop {} {\bibfield  {journal}
  {\bibinfo  {journal} {IBM journal of research and development}\ }\textbf
  {\bibinfo {volume} {5}},\ \bibinfo {pages} {183} (\bibinfo {year}
  {1961})}\BibitemShut {NoStop}%
\bibitem [{\citenamefont {Deffner}\ \emph {et~al.}(2016)\citenamefont
  {Deffner}, \citenamefont {Paz},\ and\ \citenamefont
  {Zurek}}]{deffner2016quantum}%
  \BibitemOpen
  \bibfield  {author} {\bibinfo {author} {\bibfnamefont {S.}~\bibnamefont
  {Deffner}}, \bibinfo {author} {\bibfnamefont {J.~P.}\ \bibnamefont {Paz}},\
  and\ \bibinfo {author} {\bibfnamefont {W.~H.}\ \bibnamefont {Zurek}},\
  }\bibfield  {title} {\bibinfo {title} {Quantum work and the thermodynamic
  cost of quantum measurements},\ }\href@noop {} {\bibfield  {journal}
  {\bibinfo  {journal} {Physical Review E}\ }\textbf {\bibinfo {volume} {94}},\
  \bibinfo {pages} {010103} (\bibinfo {year} {2016})}\BibitemShut {NoStop}%
\bibitem [{\citenamefont {Troiani}\ and\ \citenamefont
  {Paris}(2018)}]{troiani2018universal}%
  \BibitemOpen
  \bibfield  {author} {\bibinfo {author} {\bibfnamefont {F.}~\bibnamefont
  {Troiani}}\ and\ \bibinfo {author} {\bibfnamefont {M.~G.}\ \bibnamefont
  {Paris}},\ }\bibfield  {title} {\bibinfo {title} {Universal quantum
  magnetometry with spin states at equilibrium},\ }\href@noop {} {\bibfield
  {journal} {\bibinfo  {journal} {Physical review letters}\ }\textbf {\bibinfo
  {volume} {120}},\ \bibinfo {pages} {260503} (\bibinfo {year}
  {2018})}\BibitemShut {NoStop}%
\bibitem [{\citenamefont {Chin}\ \emph {et~al.}(2012)\citenamefont {Chin},
  \citenamefont {Huelga},\ and\ \citenamefont {Plenio}}]{chin2012quantum}%
  \BibitemOpen
  \bibfield  {author} {\bibinfo {author} {\bibfnamefont {A.~W.}\ \bibnamefont
  {Chin}}, \bibinfo {author} {\bibfnamefont {S.~F.}\ \bibnamefont {Huelga}},\
  and\ \bibinfo {author} {\bibfnamefont {M.~B.}\ \bibnamefont {Plenio}},\
  }\bibfield  {title} {\bibinfo {title} {Quantum metrology in non-markovian
  environments},\ }\href@noop {} {\bibfield  {journal} {\bibinfo  {journal}
  {Physical review letters}\ }\textbf {\bibinfo {volume} {109}},\ \bibinfo
  {pages} {233601} (\bibinfo {year} {2012})}\BibitemShut {NoStop}%
\bibitem [{\citenamefont {Fang}\ \emph {et~al.}(2019)\citenamefont {Fang},
  \citenamefont {Kruse}, \citenamefont {Lu},\ and\ \citenamefont
  {Wang}}]{fang2019nonequilibrium}%
  \BibitemOpen
  \bibfield  {author} {\bibinfo {author} {\bibfnamefont {X.}~\bibnamefont
  {Fang}}, \bibinfo {author} {\bibfnamefont {K.}~\bibnamefont {Kruse}},
  \bibinfo {author} {\bibfnamefont {T.}~\bibnamefont {Lu}},\ and\ \bibinfo
  {author} {\bibfnamefont {J.}~\bibnamefont {Wang}},\ }\bibfield  {title}
  {\bibinfo {title} {Nonequilibrium physics in biology},\ }\href@noop {}
  {\bibfield  {journal} {\bibinfo  {journal} {Reviews of Modern Physics}\
  }\textbf {\bibinfo {volume} {91}},\ \bibinfo {pages} {045004} (\bibinfo
  {year} {2019})}\BibitemShut {NoStop}%
\bibitem [{\citenamefont {De~Groot}\ and\ \citenamefont
  {Mazur}(2013)}]{de2013non}%
  \BibitemOpen
  \bibfield  {author} {\bibinfo {author} {\bibfnamefont {S.~R.}\ \bibnamefont
  {De~Groot}}\ and\ \bibinfo {author} {\bibfnamefont {P.}~\bibnamefont
  {Mazur}},\ }\href@noop {} {\emph {\bibinfo {title} {Non-equilibrium
  thermodynamics}}}\ (\bibinfo  {publisher} {Courier Corporation},\ \bibinfo
  {year} {2013})\BibitemShut {NoStop}%
\bibitem [{\citenamefont {Wang}\ \emph {et~al.}(2019)\citenamefont {Wang},
  \citenamefont {Wu},\ and\ \citenamefont {Wang}}]{wang2019steady}%
  \BibitemOpen
  \bibfield  {author} {\bibinfo {author} {\bibfnamefont {Z.}~\bibnamefont
  {Wang}}, \bibinfo {author} {\bibfnamefont {W.}~\bibnamefont {Wu}},\ and\
  \bibinfo {author} {\bibfnamefont {J.}~\bibnamefont {Wang}},\ }\bibfield
  {title} {\bibinfo {title} {Steady-state entanglement and coherence of two
  coupled qubits in equilibrium and nonequilibrium environments},\ }\href@noop
  {} {\bibfield  {journal} {\bibinfo  {journal} {Physical Review A}\ }\textbf
  {\bibinfo {volume} {99}},\ \bibinfo {pages} {042320} (\bibinfo {year}
  {2019})}\BibitemShut {NoStop}%
\bibitem [{\citenamefont {Wang}\ and\ \citenamefont
  {Wang}(2019)}]{wang2019nonequilibrium}%
  \BibitemOpen
  \bibfield  {author} {\bibinfo {author} {\bibfnamefont {X.}~\bibnamefont
  {Wang}}\ and\ \bibinfo {author} {\bibfnamefont {J.}~\bibnamefont {Wang}},\
  }\bibfield  {title} {\bibinfo {title} {Nonequilibrium effects on quantum
  correlations: Discord, mutual information, and entanglement of a
  two-fermionic system in bosonic and fermionic environments},\ }\href@noop {}
  {\bibfield  {journal} {\bibinfo  {journal} {Physical Review A}\ }\textbf
  {\bibinfo {volume} {100}},\ \bibinfo {pages} {052331} (\bibinfo {year}
  {2019})}\BibitemShut {NoStop}%
\bibitem [{\citenamefont {Zhang}\ and\ \citenamefont
  {Wang}(2014)}]{zhang2014curl}%
  \BibitemOpen
  \bibfield  {author} {\bibinfo {author} {\bibfnamefont {Z.}~\bibnamefont
  {Zhang}}\ and\ \bibinfo {author} {\bibfnamefont {J.}~\bibnamefont {Wang}},\
  }\bibfield  {title} {\bibinfo {title} {Curl flux, coherence, and population
  landscape of molecular systems: Nonequilibrium quantum steady state, energy
  (charge) transport, and thermodynamics},\ }\href@noop {} {\bibfield
  {journal} {\bibinfo  {journal} {The Journal of chemical physics}\ }\textbf
  {\bibinfo {volume} {140}},\ \bibinfo {pages} {06B622\_1} (\bibinfo {year}
  {2014})}\BibitemShut {NoStop}%
\bibitem [{\citenamefont {Li}\ \emph {et~al.}(2015)\citenamefont {Li},
  \citenamefont {Cai},\ and\ \citenamefont {Sun}}]{li2015steady}%
  \BibitemOpen
  \bibfield  {author} {\bibinfo {author} {\bibfnamefont {S.-W.}\ \bibnamefont
  {Li}}, \bibinfo {author} {\bibfnamefont {C.}~\bibnamefont {Cai}},\ and\
  \bibinfo {author} {\bibfnamefont {C.}~\bibnamefont {Sun}},\ }\bibfield
  {title} {\bibinfo {title} {Steady quantum coherence in non-equilibrium
  environment},\ }\href@noop {} {\bibfield  {journal} {\bibinfo  {journal}
  {Annals of Physics}\ }\textbf {\bibinfo {volume} {360}},\ \bibinfo {pages}
  {19} (\bibinfo {year} {2015})}\BibitemShut {NoStop}%
\bibitem [{\citenamefont {Zhang}\ \emph {et~al.}(2020)\citenamefont {Zhang},
  \citenamefont {Wu},\ and\ \citenamefont {Wang}}]{zhang2020influence}%
  \BibitemOpen
  \bibfield  {author} {\bibinfo {author} {\bibfnamefont {K.}~\bibnamefont
  {Zhang}}, \bibinfo {author} {\bibfnamefont {W.}~\bibnamefont {Wu}},\ and\
  \bibinfo {author} {\bibfnamefont {J.}~\bibnamefont {Wang}},\ }\bibfield
  {title} {\bibinfo {title} {Influence of equilibrium and nonequilibrium
  environments on macroscopic realism through the leggett-garg inequalities},\
  }\href@noop {} {\bibfield  {journal} {\bibinfo  {journal} {Physical Review
  A}\ }\textbf {\bibinfo {volume} {101}},\ \bibinfo {pages} {052334} (\bibinfo
  {year} {2020})}\BibitemShut {NoStop}%
\bibitem [{\citenamefont {Zhang}\ and\ \citenamefont
  {Wang}(2021)}]{zhang2021entanglement}%
  \BibitemOpen
  \bibfield  {author} {\bibinfo {author} {\bibfnamefont {K.}~\bibnamefont
  {Zhang}}\ and\ \bibinfo {author} {\bibfnamefont {J.}~\bibnamefont {Wang}},\
  }\bibfield  {title} {\bibinfo {title} {Entanglement versus bell nonlocality
  of quantum nonequilibrium steady states},\ }\href@noop {} {\bibfield
  {journal} {\bibinfo  {journal} {Quantum Information Processing}\ }\textbf
  {\bibinfo {volume} {20}},\ \bibinfo {pages} {1} (\bibinfo {year}
  {2021})}\BibitemShut {NoStop}%
\bibitem [{\citenamefont {Brask}\ \emph {et~al.}(2015)\citenamefont {Brask},
  \citenamefont {Haack}, \citenamefont {Brunner},\ and\ \citenamefont
  {Huber}}]{brask2015autonomous}%
  \BibitemOpen
  \bibfield  {author} {\bibinfo {author} {\bibfnamefont {J.~B.}\ \bibnamefont
  {Brask}}, \bibinfo {author} {\bibfnamefont {G.}~\bibnamefont {Haack}},
  \bibinfo {author} {\bibfnamefont {N.}~\bibnamefont {Brunner}},\ and\ \bibinfo
  {author} {\bibfnamefont {M.}~\bibnamefont {Huber}},\ }\bibfield  {title}
  {\bibinfo {title} {Autonomous quantum thermal machine for generating
  steady-state entanglement},\ }\href@noop {} {\bibfield  {journal} {\bibinfo
  {journal} {New Journal of Physics}\ }\textbf {\bibinfo {volume} {17}},\
  \bibinfo {pages} {113029} (\bibinfo {year} {2015})}\BibitemShut {NoStop}%
\bibitem [{\citenamefont {Tacchino}\ \emph {et~al.}(2018)\citenamefont
  {Tacchino}, \citenamefont {Auff{\`e}ves}, \citenamefont {Santos},\ and\
  \citenamefont {Gerace}}]{tacchino2018steady}%
  \BibitemOpen
  \bibfield  {author} {\bibinfo {author} {\bibfnamefont {F.}~\bibnamefont
  {Tacchino}}, \bibinfo {author} {\bibfnamefont {A.}~\bibnamefont
  {Auff{\`e}ves}}, \bibinfo {author} {\bibfnamefont {M.}~\bibnamefont
  {Santos}},\ and\ \bibinfo {author} {\bibfnamefont {D.}~\bibnamefont
  {Gerace}},\ }\bibfield  {title} {\bibinfo {title} {Steady state entanglement
  beyond thermal limits},\ }\href@noop {} {\bibfield  {journal} {\bibinfo
  {journal} {Physical review letters}\ }\textbf {\bibinfo {volume} {120}},\
  \bibinfo {pages} {063604} (\bibinfo {year} {2018})}\BibitemShut {NoStop}%
\bibitem [{\citenamefont {Brunner}\ \emph {et~al.}(2014)\citenamefont
  {Brunner}, \citenamefont {Huber}, \citenamefont {Linden}, \citenamefont
  {Popescu}, \citenamefont {Silva},\ and\ \citenamefont
  {Skrzypczyk}}]{brunner2014entanglement}%
  \BibitemOpen
  \bibfield  {author} {\bibinfo {author} {\bibfnamefont {N.}~\bibnamefont
  {Brunner}}, \bibinfo {author} {\bibfnamefont {M.}~\bibnamefont {Huber}},
  \bibinfo {author} {\bibfnamefont {N.}~\bibnamefont {Linden}}, \bibinfo
  {author} {\bibfnamefont {S.}~\bibnamefont {Popescu}}, \bibinfo {author}
  {\bibfnamefont {R.}~\bibnamefont {Silva}},\ and\ \bibinfo {author}
  {\bibfnamefont {P.}~\bibnamefont {Skrzypczyk}},\ }\bibfield  {title}
  {\bibinfo {title} {Entanglement enhances cooling in microscopic quantum
  refrigerators},\ }\href@noop {} {\bibfield  {journal} {\bibinfo  {journal}
  {Physical Review E}\ }\textbf {\bibinfo {volume} {89}},\ \bibinfo {pages}
  {032115} (\bibinfo {year} {2014})}\BibitemShut {NoStop}%
\bibitem [{\citenamefont {Manzano}\ \emph {et~al.}(2012)\citenamefont
  {Manzano}, \citenamefont {Tiersch}, \citenamefont {Asadian},\ and\
  \citenamefont {Briegel}}]{manzano2012quantum}%
  \BibitemOpen
  \bibfield  {author} {\bibinfo {author} {\bibfnamefont {D.}~\bibnamefont
  {Manzano}}, \bibinfo {author} {\bibfnamefont {M.}~\bibnamefont {Tiersch}},
  \bibinfo {author} {\bibfnamefont {A.}~\bibnamefont {Asadian}},\ and\ \bibinfo
  {author} {\bibfnamefont {H.~J.}\ \bibnamefont {Briegel}},\ }\bibfield
  {title} {\bibinfo {title} {Quantum transport efficiency and fourier's law},\
  }\href@noop {} {\bibfield  {journal} {\bibinfo  {journal} {Physical Review
  E}\ }\textbf {\bibinfo {volume} {86}},\ \bibinfo {pages} {061118} (\bibinfo
  {year} {2012})}\BibitemShut {NoStop}%
\bibitem [{\citenamefont {Wang}\ \emph {et~al.}(2021)\citenamefont {Wang},
  \citenamefont {Zhang},\ and\ \citenamefont {Wang}}]{wang2021excitation}%
  \BibitemOpen
  \bibfield  {author} {\bibinfo {author} {\bibfnamefont {X.}~\bibnamefont
  {Wang}}, \bibinfo {author} {\bibfnamefont {Z.}~\bibnamefont {Zhang}},\ and\
  \bibinfo {author} {\bibfnamefont {J.}~\bibnamefont {Wang}},\ }\bibfield
  {title} {\bibinfo {title} {Excitation-energy transfer under strong laser
  drive},\ }\href@noop {} {\bibfield  {journal} {\bibinfo  {journal} {Physical
  Review A}\ }\textbf {\bibinfo {volume} {103}},\ \bibinfo {pages} {013516}
  (\bibinfo {year} {2021})}\BibitemShut {NoStop}%
\bibitem [{\citenamefont {Onuchic}\ and\ \citenamefont
  {Wolynes}(1988)}]{onuchic1988classical}%
  \BibitemOpen
  \bibfield  {author} {\bibinfo {author} {\bibfnamefont {J.~N.}\ \bibnamefont
  {Onuchic}}\ and\ \bibinfo {author} {\bibfnamefont {P.~G.}\ \bibnamefont
  {Wolynes}},\ }\bibfield  {title} {\bibinfo {title} {Classical and quantum
  pictures of reaction dynamics in condensed matter: Resonances, dephasing, and
  all that},\ }\href@noop {} {\bibfield  {journal} {\bibinfo  {journal} {The
  Journal of Physical Chemistry}\ }\textbf {\bibinfo {volume} {92}},\ \bibinfo
  {pages} {6495} (\bibinfo {year} {1988})}\BibitemShut {NoStop}%
\bibitem [{\citenamefont {Sinaysky}\ \emph {et~al.}(2008)\citenamefont
  {Sinaysky}, \citenamefont {Petruccione},\ and\ \citenamefont
  {Burgarth}}]{sinaysky2008dynamics}%
  \BibitemOpen
  \bibfield  {author} {\bibinfo {author} {\bibfnamefont {I.}~\bibnamefont
  {Sinaysky}}, \bibinfo {author} {\bibfnamefont {F.}~\bibnamefont
  {Petruccione}},\ and\ \bibinfo {author} {\bibfnamefont {D.}~\bibnamefont
  {Burgarth}},\ }\bibfield  {title} {\bibinfo {title} {Dynamics of
  nonequilibrium thermal entanglement},\ }\href@noop {} {\bibfield  {journal}
  {\bibinfo  {journal} {Physical Review A}\ }\textbf {\bibinfo {volume} {78}},\
  \bibinfo {pages} {062301} (\bibinfo {year} {2008})}\BibitemShut {NoStop}%
\bibitem [{\citenamefont {Gunlycke}\ \emph {et~al.}(2001)\citenamefont
  {Gunlycke}, \citenamefont {Kendon}, \citenamefont {Vedral},\ and\
  \citenamefont {Bose}}]{gunlycke2001thermal}%
  \BibitemOpen
  \bibfield  {author} {\bibinfo {author} {\bibfnamefont {D.}~\bibnamefont
  {Gunlycke}}, \bibinfo {author} {\bibfnamefont {V.}~\bibnamefont {Kendon}},
  \bibinfo {author} {\bibfnamefont {V.}~\bibnamefont {Vedral}},\ and\ \bibinfo
  {author} {\bibfnamefont {S.}~\bibnamefont {Bose}},\ }\bibfield  {title}
  {\bibinfo {title} {Thermal concurrence mixing in a one-dimensional ising
  model},\ }\href@noop {} {\bibfield  {journal} {\bibinfo  {journal} {Physical
  Review A}\ }\textbf {\bibinfo {volume} {64}},\ \bibinfo {pages} {042302}
  (\bibinfo {year} {2001})}\BibitemShut {NoStop}%
\bibitem [{\citenamefont {Petrosyan}\ and\ \citenamefont
  {Kurizki}(2002)}]{petrosyan2002scalable}%
  \BibitemOpen
  \bibfield  {author} {\bibinfo {author} {\bibfnamefont {D.}~\bibnamefont
  {Petrosyan}}\ and\ \bibinfo {author} {\bibfnamefont {G.}~\bibnamefont
  {Kurizki}},\ }\bibfield  {title} {\bibinfo {title} {Scalable solid-state
  quantum processor using subradiant two-atom states},\ }\href@noop {}
  {\bibfield  {journal} {\bibinfo  {journal} {Physical review letters}\
  }\textbf {\bibinfo {volume} {89}},\ \bibinfo {pages} {207902} (\bibinfo
  {year} {2002})}\BibitemShut {NoStop}%
\bibitem [{\citenamefont {Petrosyan}\ and\ \citenamefont
  {Kurizki}(2006)}]{petrosyan2006quantum}%
  \BibitemOpen
  \bibfield  {author} {\bibinfo {author} {\bibfnamefont {D.}~\bibnamefont
  {Petrosyan}}\ and\ \bibinfo {author} {\bibfnamefont {G.}~\bibnamefont
  {Kurizki}},\ }\bibfield  {title} {\bibinfo {title} {Quantum computer with
  dipole-dipole interacting two-level systems},\ }\href@noop {} {\bibfield
  {journal} {\bibinfo  {journal} {Quantum Information \& Computation}\ }\textbf
  {\bibinfo {volume} {6}},\ \bibinfo {pages} {1} (\bibinfo {year}
  {2006})}\BibitemShut {NoStop}%
\bibitem [{\citenamefont {Joachim}\ and\ \citenamefont
  {Ratner}(2005)}]{joachim2005molecular}%
  \BibitemOpen
  \bibfield  {author} {\bibinfo {author} {\bibfnamefont {C.}~\bibnamefont
  {Joachim}}\ and\ \bibinfo {author} {\bibfnamefont {M.~A.}\ \bibnamefont
  {Ratner}},\ }\bibfield  {title} {\bibinfo {title} {Molecular electronics:
  Some views on transport junctions and beyond},\ }\href@noop {} {\bibfield
  {journal} {\bibinfo  {journal} {Proceedings of the National Academy of
  Sciences}\ }\textbf {\bibinfo {volume} {102}},\ \bibinfo {pages} {8801}
  (\bibinfo {year} {2005})}\BibitemShut {NoStop}%
\bibitem [{\citenamefont {Galperin}\ \emph {et~al.}(2005)\citenamefont
  {Galperin}, \citenamefont {Ratner},\ and\ \citenamefont
  {Nitzan}}]{galperin2005hysteresis}%
  \BibitemOpen
  \bibfield  {author} {\bibinfo {author} {\bibfnamefont {M.}~\bibnamefont
  {Galperin}}, \bibinfo {author} {\bibfnamefont {M.~A.}\ \bibnamefont
  {Ratner}},\ and\ \bibinfo {author} {\bibfnamefont {A.}~\bibnamefont
  {Nitzan}},\ }\bibfield  {title} {\bibinfo {title} {Hysteresis, switching, and
  negative differential resistance in molecular junctions: A polaron model},\
  }\href@noop {} {\bibfield  {journal} {\bibinfo  {journal} {Nano letters}\
  }\textbf {\bibinfo {volume} {5}},\ \bibinfo {pages} {125} (\bibinfo {year}
  {2005})}\BibitemShut {NoStop}%
\bibitem [{\citenamefont {Pezz{\`e}}\ \emph {et~al.}(2018)\citenamefont
  {Pezz{\`e}}, \citenamefont {Smerzi}, \citenamefont {Oberthaler},
  \citenamefont {Schmied},\ and\ \citenamefont {Treutlein}}]{pezze2018quantum}%
  \BibitemOpen
  \bibfield  {author} {\bibinfo {author} {\bibfnamefont {L.}~\bibnamefont
  {Pezz{\`e}}}, \bibinfo {author} {\bibfnamefont {A.}~\bibnamefont {Smerzi}},
  \bibinfo {author} {\bibfnamefont {M.~K.}\ \bibnamefont {Oberthaler}},
  \bibinfo {author} {\bibfnamefont {R.}~\bibnamefont {Schmied}},\ and\ \bibinfo
  {author} {\bibfnamefont {P.}~\bibnamefont {Treutlein}},\ }\bibfield  {title}
  {\bibinfo {title} {Quantum metrology with nonclassical states of atomic
  ensembles},\ }\href@noop {} {\bibfield  {journal} {\bibinfo  {journal}
  {Reviews of Modern Physics}\ }\textbf {\bibinfo {volume} {90}},\ \bibinfo
  {pages} {035005} (\bibinfo {year} {2018})}\BibitemShut {NoStop}%
\bibitem [{\citenamefont {Lu}\ \emph {et~al.}(2013)\citenamefont {Lu},
  \citenamefont {Sun}, \citenamefont {Wang}, \citenamefont {Luo},\ and\
  \citenamefont {Oh}}]{lu2013broadcasting}%
  \BibitemOpen
  \bibfield  {author} {\bibinfo {author} {\bibfnamefont {X.-M.}\ \bibnamefont
  {Lu}}, \bibinfo {author} {\bibfnamefont {Z.}~\bibnamefont {Sun}}, \bibinfo
  {author} {\bibfnamefont {X.}~\bibnamefont {Wang}}, \bibinfo {author}
  {\bibfnamefont {S.}~\bibnamefont {Luo}},\ and\ \bibinfo {author}
  {\bibfnamefont {C.}~\bibnamefont {Oh}},\ }\bibfield  {title} {\bibinfo
  {title} {Broadcasting quantum fisher information},\ }\href@noop {} {\bibfield
   {journal} {\bibinfo  {journal} {Physical Review A}\ }\textbf {\bibinfo
  {volume} {87}},\ \bibinfo {pages} {050302} (\bibinfo {year}
  {2013})}\BibitemShut {NoStop}%
\bibitem [{\citenamefont {Hill}\ and\ \citenamefont
  {Wootters}(1997)}]{hill1997entanglement}%
  \BibitemOpen
  \bibfield  {author} {\bibinfo {author} {\bibfnamefont {S.}~\bibnamefont
  {Hill}}\ and\ \bibinfo {author} {\bibfnamefont {W.~K.}\ \bibnamefont
  {Wootters}},\ }\bibfield  {title} {\bibinfo {title} {Entanglement of a pair
  of quantum bits},\ }\href@noop {} {\bibfield  {journal} {\bibinfo  {journal}
  {Physical review letters}\ }\textbf {\bibinfo {volume} {78}},\ \bibinfo
  {pages} {5022} (\bibinfo {year} {1997})}\BibitemShut {NoStop}%
\bibitem [{\citenamefont {Wootters}(1998)}]{wootters1998entanglement}%
  \BibitemOpen
  \bibfield  {author} {\bibinfo {author} {\bibfnamefont {W.~K.}\ \bibnamefont
  {Wootters}},\ }\bibfield  {title} {\bibinfo {title} {Entanglement of
  formation of an arbitrary state of two qubits},\ }\href@noop {} {\bibfield
  {journal} {\bibinfo  {journal} {Physical Review Letters}\ }\textbf {\bibinfo
  {volume} {80}},\ \bibinfo {pages} {2245} (\bibinfo {year}
  {1998})}\BibitemShut {NoStop}%
\bibitem [{\citenamefont {Jaeger}(2007)}]{jaeger2007quantum}%
  \BibitemOpen
  \bibfield  {author} {\bibinfo {author} {\bibfnamefont {G.}~\bibnamefont
  {Jaeger}},\ }\href@noop {} {\emph {\bibinfo {title} {Quantum information}}}\
  (\bibinfo  {publisher} {Springer},\ \bibinfo {year} {2007})\BibitemShut
  {NoStop}%
\bibitem [{\citenamefont {Hofer}\ \emph {et~al.}(2017)\citenamefont {Hofer},
  \citenamefont {Perarnau-Llobet}, \citenamefont {Miranda}, \citenamefont
  {Haack}, \citenamefont {Silva}, \citenamefont {Brask},\ and\ \citenamefont
  {Brunner}}]{hofer2017markovian}%
  \BibitemOpen
  \bibfield  {author} {\bibinfo {author} {\bibfnamefont {P.~P.}\ \bibnamefont
  {Hofer}}, \bibinfo {author} {\bibfnamefont {M.}~\bibnamefont
  {Perarnau-Llobet}}, \bibinfo {author} {\bibfnamefont {L.~D.~M.}\ \bibnamefont
  {Miranda}}, \bibinfo {author} {\bibfnamefont {G.}~\bibnamefont {Haack}},
  \bibinfo {author} {\bibfnamefont {R.}~\bibnamefont {Silva}}, \bibinfo
  {author} {\bibfnamefont {J.~B.}\ \bibnamefont {Brask}},\ and\ \bibinfo
  {author} {\bibfnamefont {N.}~\bibnamefont {Brunner}},\ }\bibfield  {title}
  {\bibinfo {title} {Markovian master equations for quantum thermal machines:
  local versus global approach},\ }\href@noop {} {\bibfield  {journal}
  {\bibinfo  {journal} {New Journal of Physics}\ }\textbf {\bibinfo {volume}
  {19}},\ \bibinfo {pages} {123037} (\bibinfo {year} {2017})}\BibitemShut
  {NoStop}%
\bibitem [{\citenamefont {Gonz{\'a}lez}\ \emph {et~al.}(2017)\citenamefont
  {Gonz{\'a}lez}, \citenamefont {Correa}, \citenamefont {Nocerino},
  \citenamefont {Palao}, \citenamefont {Alonso},\ and\ \citenamefont
  {Adesso}}]{gonzalez2017testing}%
  \BibitemOpen
  \bibfield  {author} {\bibinfo {author} {\bibfnamefont {J.~O.}\ \bibnamefont
  {Gonz{\'a}lez}}, \bibinfo {author} {\bibfnamefont {L.~A.}\ \bibnamefont
  {Correa}}, \bibinfo {author} {\bibfnamefont {G.}~\bibnamefont {Nocerino}},
  \bibinfo {author} {\bibfnamefont {J.~P.}\ \bibnamefont {Palao}}, \bibinfo
  {author} {\bibfnamefont {D.}~\bibnamefont {Alonso}},\ and\ \bibinfo {author}
  {\bibfnamefont {G.}~\bibnamefont {Adesso}},\ }\bibfield  {title} {\bibinfo
  {title} {Testing the validity of the ‘local’and ‘global’gkls master
  equations on an exactly solvable model},\ }\href@noop {} {\bibfield
  {journal} {\bibinfo  {journal} {Open Systems \& Information Dynamics}\
  }\textbf {\bibinfo {volume} {24}},\ \bibinfo {pages} {1740010} (\bibinfo
  {year} {2017})}\BibitemShut {NoStop}%
\bibitem [{\citenamefont {Carmichael}\ and\ \citenamefont
  {Walls}(1973)}]{carmichael1973master}%
  \BibitemOpen
  \bibfield  {author} {\bibinfo {author} {\bibfnamefont {H.}~\bibnamefont
  {Carmichael}}\ and\ \bibinfo {author} {\bibfnamefont {D.}~\bibnamefont
  {Walls}},\ }\bibfield  {title} {\bibinfo {title} {Master equation for
  strongly interacting systems},\ }\href@noop {} {\bibfield  {journal}
  {\bibinfo  {journal} {Journal of Physics A: Mathematical, Nuclear and
  General}\ }\textbf {\bibinfo {volume} {6}},\ \bibinfo {pages} {1552}
  (\bibinfo {year} {1973})}\BibitemShut {NoStop}%
\bibitem [{\citenamefont {Knysh}\ \emph {et~al.}(2011)\citenamefont {Knysh},
  \citenamefont {Smelyanskiy},\ and\ \citenamefont
  {Durkin}}]{knysh2011scaling}%
  \BibitemOpen
  \bibfield  {author} {\bibinfo {author} {\bibfnamefont {S.}~\bibnamefont
  {Knysh}}, \bibinfo {author} {\bibfnamefont {V.~N.}\ \bibnamefont
  {Smelyanskiy}},\ and\ \bibinfo {author} {\bibfnamefont {G.~A.}\ \bibnamefont
  {Durkin}},\ }\bibfield  {title} {\bibinfo {title} {Scaling laws for precision
  in quantum interferometry and the bifurcation landscape of the optimal
  state},\ }\href@noop {} {\bibfield  {journal} {\bibinfo  {journal} {Physical
  Review A}\ }\textbf {\bibinfo {volume} {83}},\ \bibinfo {pages} {021804}
  (\bibinfo {year} {2011})}\BibitemShut {NoStop}%
\bibitem [{\citenamefont {Levy}\ and\ \citenamefont
  {Kosloff}(2014)}]{levy2014local}%
  \BibitemOpen
  \bibfield  {author} {\bibinfo {author} {\bibfnamefont {A.}~\bibnamefont
  {Levy}}\ and\ \bibinfo {author} {\bibfnamefont {R.}~\bibnamefont {Kosloff}},\
  }\bibfield  {title} {\bibinfo {title} {The local approach to quantum
  transport may violate the second law of thermodynamics},\ }\href@noop {}
  {\bibfield  {journal} {\bibinfo  {journal} {EPL (Europhysics Letters)}\
  }\textbf {\bibinfo {volume} {107}},\ \bibinfo {pages} {20004} (\bibinfo
  {year} {2014})}\BibitemShut {NoStop}%
\bibitem [{\citenamefont {Landi}\ and\ \citenamefont
  {Paternostro}(2020)}]{landi2020irreversible}%
  \BibitemOpen
  \bibfield  {author} {\bibinfo {author} {\bibfnamefont {G.~T.}\ \bibnamefont
  {Landi}}\ and\ \bibinfo {author} {\bibfnamefont {M.}~\bibnamefont
  {Paternostro}},\ }\bibfield  {title} {\bibinfo {title} {Irreversible entropy
  production, from quantum to classical},\ }\href@noop {} {\bibfield  {journal}
  {\bibinfo  {journal} {arXiv preprint arXiv:2009.07668}\ } (\bibinfo {year}
  {2020})}\BibitemShut {NoStop}%
\bibitem [{\citenamefont {Jang}\ \emph {et~al.}(2004)\citenamefont {Jang},
  \citenamefont {Newton},\ and\ \citenamefont
  {Silbey}}]{jang2004multichromophoric}%
  \BibitemOpen
  \bibfield  {author} {\bibinfo {author} {\bibfnamefont {S.}~\bibnamefont
  {Jang}}, \bibinfo {author} {\bibfnamefont {M.~D.}\ \bibnamefont {Newton}},\
  and\ \bibinfo {author} {\bibfnamefont {R.~J.}\ \bibnamefont {Silbey}},\
  }\bibfield  {title} {\bibinfo {title} {Multichromophoric f{\"o}rster
  resonance energy transfer},\ }\href@noop {} {\bibfield  {journal} {\bibinfo
  {journal} {Physical review letters}\ }\textbf {\bibinfo {volume} {92}},\
  \bibinfo {pages} {218301} (\bibinfo {year} {2004})}\BibitemShut {NoStop}%
\bibitem [{\citenamefont {Clapp}\ \emph {et~al.}(2006)\citenamefont {Clapp},
  \citenamefont {Medintz},\ and\ \citenamefont {Mattoussi}}]{clapp2006forster}%
  \BibitemOpen
  \bibfield  {author} {\bibinfo {author} {\bibfnamefont {A.~R.}\ \bibnamefont
  {Clapp}}, \bibinfo {author} {\bibfnamefont {I.~L.}\ \bibnamefont {Medintz}},\
  and\ \bibinfo {author} {\bibfnamefont {H.}~\bibnamefont {Mattoussi}},\
  }\bibfield  {title} {\bibinfo {title} {F{\"o}rster resonance energy transfer
  investigations using quantum-dot fluorophores},\ }\href@noop {} {\bibfield
  {journal} {\bibinfo  {journal} {ChemPhysChem}\ }\textbf {\bibinfo {volume}
  {7}},\ \bibinfo {pages} {47} (\bibinfo {year} {2006})}\BibitemShut {NoStop}%
\bibitem [{\citenamefont {Yu}\ and\ \citenamefont
  {Eberly}(2009)}]{yu2009sudden}%
  \BibitemOpen
  \bibfield  {author} {\bibinfo {author} {\bibfnamefont {T.}~\bibnamefont
  {Yu}}\ and\ \bibinfo {author} {\bibfnamefont {J.}~\bibnamefont {Eberly}},\
  }\bibfield  {title} {\bibinfo {title} {Sudden death of entanglement},\
  }\href@noop {} {\bibfield  {journal} {\bibinfo  {journal} {Science}\ }\textbf
  {\bibinfo {volume} {323}},\ \bibinfo {pages} {598} (\bibinfo {year}
  {2009})}\BibitemShut {NoStop}%
\end{thebibliography}
\end{document}